\documentclass[universe,article,accept,pdftex,moreauthors]{Definitions/mdpi} 

\firstpage{1} 
\pubvolume{1}
\issuenum{1}
\articlenumber{0}
\pubyear{2024}
\copyrightyear{2024}
\externaleditor{Academic Editor(s):  {Alexandre Marcowith}} 
\datereceived{29 April 2024} 
\daterevised{18 June 2024} 
\dateaccepted{25 June 2024} 
\datepublished{ } 
\hreflink{https://doi.org/} 

\usepackage{graphicx}%
\usepackage{multirow}%
\usepackage{amsmath,amssymb,amsfonts}%
\usepackage{amsthm}%
\usepackage{mathrsfs}%
\usepackage[title]{appendix}%
\usepackage{textcomp}%
\usepackage{manyfoot}%
\usepackage{booktabs}%
\usepackage{listings}%
\usepackage[export]{adjustbox}
\usepackage{subcaption}
\usepackage[font=small]{subcaption}
\usepackage[version=4]{mhchem} 

\Title{Molecular Formation in Low-Metallicity Hot Cores}

\TitleCitation{Molecular Formation in Low-Metallicity Hot Cores}

\Author{{Yara Sobhy} $^{1,}$*\orcidA{}, Hideko Nomura $^{2,3}$, Tetsuo Yamamoto $^{4}$ and Osama Shalabeia $^{1,5}$} 

\AuthorNames{Yara Sobhy, Hideko Nomura, Tetsuo Yamamoto and Osama Shalabeia}

\AuthorCitation{Sobhy, Y.; Nomura, H.; Yamamoto, T.; Shalabeia, O.}

\address{%
$^{1}$ \quad Department of Astronomy, Space Science and Meteorology, Faculty of Science, Cairo University,  \mbox{Giza 12613, Egypt}; {shalabiea@sci.cu.edu.eg}\\
$^{2}$ \quad Division of Science, National Astronomical Observatory of Japan, 2-21-1 Osawa, \mbox{Mitaka, {Tokyo}, {181-8588}, Japan; {hideko.nomura@nao.ac.jp}}\\ 
$^{3}$ \quad Astronomical Science Program, The Graduate University for Advanced Studies, SOKENDAI, 2-21-1 Osawa, \mbox{Mitaka, {Tokyo}, {181-8588}, Japan}\\
$^{4}$ \quad Institute of Low Temperature Science, Hokkaido University, \mbox{Sapporo {060-0819, Hokkaido}, Japan}\\
$^{5}$ \quad Faculty of Navigation Science and Space Technonolgy, Beni-Suef University, \mbox{{Beni-Suef}, 62521, Egypt}}

\corres{Correspondence: ymsobhy@sci.cu.edu.eg }

\abstract{The chemical complexity in low-metallicity hot cores has been confirmed by observations. We investigate the effect of varying physical parameters, such as temperature, density and the cosmic ray ionisation rate (CRIR), on the molecular abundance evolution in low-metallicity hot cores using the UMIST gas phase chemical model. CRIR had the strongest effect on molecular abundance. The resultant molecular abundances were divided into three categories with different trends in time evolution.~We compared our results with the observations of hot cores in the Large Magellanic Cloud (LMC). Our model fits best with the observations at a time of around $10^5$ years after the evaporation of ice and at the CRIR of $1.36 \times 10^{-16}$ s$^{-1}$. The resultant abundances of the oxygen-bearing complex organic molecules (COMs), such as \ce{CH3OH}, \ce{HCOOCH3} and \ce{CH3OCH3}, do not fit with observations in the same physical condition and may be located in a different physical environment. Our results suggest that investigating the CRIR value is crucial to predict the molecular evolution in LMC hot~\mbox{cores}.}


\begin{document}


\section{Introduction}\label{sec1}

Studying the chemical evolution in hot cores is crucial to understanding how the molecules evolve in star-forming regions. Hot cores are characterised by a relatively high density (can reach $10^{7}$ cm$^{-3}$) and high temperature ($\ge$100 K) near the central star that has a radius of about 0.1 pc. 
The Orion molecular hot core is one of the richest factories for different simple and complex molecules. Observations of the chemistry of this hot core (due to its \mbox{vicinity $\approx$400 pc}), progressed from resolutions that could not identify different regions in the core~\cite{Blake1987} to observations which enabled us to distinguish multiple clumps~\cite{Wright2017}.

The first hot core (ST11) in the LMC was detected by ALMA~\cite{Shimon2016}. The LMC is an irregular dwarf galaxy which is close to the Milky Way (about 50 kpc away;~\cite{Pietrz2013}), and it is characterised by its low-metallicity environment of about half of the solar metallicity~\cite{Russell1992, Dufour1982, Wester1990, Rolleston2002}. As a result of the low metallicity,that is, the low dust abundance, the UV radiation field is high, which increases the photoionisation and photodissociation processes. Thus, exploring LMC chemistry can give us clues about the early metal-poor universe.

ST11 was not the only hot core that was observed in the LMC; also, N113 A1 and B3~\cite{Sewilo2018}, ST16~\cite{Shimon2020}, N105–2 A and 2 B~\cite{Sewilo2022} were observed. Despite the low-metallicity environments in LMC hot cores, large COMs, such as methanol (\ce{CH3OH}) and methyl cyanide (\ce{CH3CN}), were detected~\cite{Sewilo2018, Shimon2020, Sewilo2022}.
\ce{CH3OH} is the parent species of larger COMs, such as dimethy ether (\ce{CH3OCH3}) and methyl formate (\ce{HCOOCH3}), and it is formed on the icy mantle of the dust-grain surfaces by the hydrogenation of CO in cold and dense prestellar cores. Then, methanol can be detected after its sublimation to the gas phase. The deficiency of methanol abundance in LMC was indicated~\cite{Acharyya2015, Shimonb2016, Shimon2016, Nishimura2016}. However, further observations reveal that \ce{CH3OH} is rich in some hot cores and even larger COMs are detected; for example, \ce{CH3OCH3} in N113 A1 and B3, N105-2 A and 2 B~\cite{Sewilo2018, Sewilo2022} in addition to \ce{HCOOCH3} in N113 A1 and B3~\cite{Sewilo2018}. Due to these variations in COM abundances, the detected LMC hot cores are classified into two categories, according to their compositions and the abundance of the organic molecules to organic-rich and organic-poor hot cores~\cite{Shimon2020, Sewilo2022}. N113 A1 and B3 as well as N105–2 A and 2 B are the organic-rich hot cores. These cores have relatively similar molecular abundances. {Ref.}~\cite{Sewilo2022} claimed that the hot core N105-2 A may contain the formamide molecule (\ce{NH2CHO}), and confirming its presence will be the first detection of an astrobiological molecule in an extragalactic low-metallicity environment. On the contrary, ST11 and ST16 are classified as organic-poor hot~cores.  

Moving on to the sulphur-bearing species, the LMC hot cores have close values of \ce{SO2}, SO and OCS abundances. However, the abundances of CS and \ce{H2CS} in the hot cores N113 A1 and B3, in addition to N105–2 A and 2 B, is higher than that in the ST16 and ST11.
A variety of nitrogen-bearing species have been detected in LMC hot cores, for example, NO, \ce{CH3CN}, \ce{HNCO} and \ce{HC3N}. NO is overabundant in ST11 and ST16, and its abundances are higher than those in galactic hot cores after correcting for metallicity~\cite{Shimon2020}.

Many previous hot core models were constructed to reproduce the observed galactic hot core chemistry. {Ref.}~\cite{Brown1988} was the first to model hot core chemistry and compare their results with the observations of the Orion hot core. Their three-phase model ended with hot core formation after the collapsing and heating of a high-density cold clump. {Ref.}~\cite{Caselli1993} simulated a dynamical--chemical model of a massive star-forming region to compare the resultant abundances from this model with the observations of the Orion hot core and the Compact Ridge. They assumed that there was a temperature and density gradient in the two regions. {Ref.}~\cite{Millar1997} constructed a model to study the three regions in the molecular cloud G34.3+0.15, which were the ultracompact core, the compact core and the halo. Each of these regions has its own temperature, density and visual extinction. They pointed out that the variation in the mantle composition with the distance from the central star should be taken into account. This was taken into consideration by~\cite{Nomura2004}, where they modelled the density and temperature profiles of the hot cores using the radiative transfer calculations for the same G34.3 cloud.

The first attempt to model the molecular evolution in the LMC was by~\cite{Millar1990}. When they performed calculations of the pseudo-time-dependent model for chemistry in the dark clouds of the LMC and the Small Magellanic Cloud (SMC), they found that the molecular abundances depended mainly on the abundances of oxygen, nitrogen and carbon, except for CO. {Ref.} \cite{Acharyya2015} studied the effect of varying different physical parameters, such as density, visual extinction and dust temperature, on the molecules in the LMC, as well as considering the low metallicity in the LMC. They found that despite the LMC's low-metallicity, it contains various molecules. Another attempt to study the chemical evolution in LMC was performed by the gas--dust model of~\cite{Pauly2018}, where they focused on the impact of varying the dust temperatures and the elemental abundances on the envelopes of young stellar objects in the magellanic clouds. \ce{CH3OH} was found to be produced efficiently in metal-poor environments of magellanic clouds.

In this paper, we aim to investigate the variation in temperature, density and CRIR on the resultant molecular abundances in LMC hot cores. Then, we discuss the resultant abundances with observations. We describe the chemistry and the physics of our model in Section~\ref{sec2}. The resultant effects of the various physical parameters are described in Section~\ref{sec3}. Our explanation and conclusion of the results are in Section~\ref{sec4} and Section~\ref{sec5}, respectively.
 
\section{Materials and Methods}\label{sec2}

To study the chemical evolution in LMC hot cores, we used the gas phase chemical network of the dark cloud model code from the fifth release accessed on 1 June 2021, which we downloaded from the UMIST Database for Astrochemistry, RATE12~\cite{Mc2013}.  
We changed the initial molecular abundances and the physical conditions to match the environment in the LMC hot cores. The reaction network of the model contained 6173 gas phase reactions and involved \linebreak 467~\mbox{species}.

\subsection{The Physical Conditions}\label{subsec2.1}

Temperature, density and CRIR affect the hot core chemistry significantly. Table~\ref{tab:phys} shows the values of the physical parameters that we used in our model. Since hot cores are formed as a result of the gravitational collapse of molecular cloud cores and the formation of massive stars at the centre of cores, which heat the surrounding materials, there are temperature and density gradients inside the hot cores, depending on their distance from the central stars. In order to cover the gradients, we set the temperature at 50~K, 100~K, 150~K and 200~K, and the density at $2 \times 10^5$ cm$^{-3}$, $2 \times 10^6$ cm$^{-3}$ and $2 \times 10^7$ cm$^{-3}$ (\mbox{{\cite{Shimon2020,Hatchell1998}}}). 
The value of the visual extinction that we used in the model was 640~\cite{Millar1997}. Therefore, photochemistry was negligible in this model since the visual extinction was large.  
The CRIR is defined as the total ionisation rate of molecular hydrogen. The value of CRIR included in UMIST code is the standard galactic value, which is $1.36 \times 10^{-17}$ s$^{-1}$. In the diffuse environment, the cosmic-ray ionisation rate can be higher (e.g.,~\cite{Padovani2024}).
According to the Fermi LAT gamma-ray map, the global average CRIR in the LMC could be lower than that in the Milky Way, as low as $\sim$$10^{-18}$ s$^{-1}$~\cite{Abdo2010, Pauly2018}. However, due to the presence of about 60 confirmed SNRs in the LMC~\cite{Maggi2016, Ou2018}, the local CRIR value in a hot core could be as high as $10^{-15}$ s$^{-1}$, when the hot core is located in the neighbourhood of SNRs~\cite{Vaupr2014}. In addition, some Galactic star-forming regions show a locally high CRIR (e.g.,~\cite{Morales2014,Cabedo2023}) (see Section~\ref{subsec4.1} for more details). Therefore, we varied the CRIR value between \mbox{$1.36 \times 10^{-18}$ s$^{-1}$} and \mbox{$1.36 \times 10^{-15}$ s$^{-1}$~\cite{Vaupr2014}} with an order of magnitude.

\begin{table}[H]%
\caption{{The} values of the physical parameters used in the model.} 
\setlength{\tabcolsep}{14.55mm}
\begin{tabular}{cc}
\toprule
\textbf{Parameter}                   &                \textbf{Values}  \\ 
\midrule
Temperature                          &               $50$ K--$200$ K \\
Density                              &               $2 \times 10^{5}$ cm$^{-3}$--$2 \times 10^{7}$ cm$^{-3}$ \\
CR Ionisation rate                   &               $1.36 \times 10^{-18}$ s$^{-1}$--$1.36 \times 10^{-15}$ s$^{-1}$ \\ 
\bottomrule
\label{tab:phys}
\end{tabular}
\end{table}

\vspace*{-21pt}

\subsection{The Initial Molecular Abundances}\label{subsec2.2}

Since the chemistry in the LMC is characterised by low metallicity, the initial molecular abundances which we used in our LMC hot core model have lower values than those in the galactic hot cores. 
We derived the low-metallicity initial molecular abundances for the LMC using the elemental abundance ratios of the following elements: carbon, oxygen, nitrogen and sulphur relative to hydrogen. These molecular abundances are calculated using the ratio between the solar elemental abundances from~\cite{Asplund2021} and the LMC elemental abundances from~\cite{Russell1992} (as shown in Table~\ref{tab:elemental abundances ratios}). This ratio, $f$, was used before by~\cite{Acharyya2015} and it is given as follows: 
\begin{equation}
    f = \frac{ \rm Solar\ Abundance\ of\ an\ element}{\rm LMC\ Abundance\ of\ the\ same\ element}.
\end{equation}
\vspace{-18pt}
\begin{table}[H]%
\caption{The solar and LMC elemental abundances and the ratio between them.}
\setlength{\tabcolsep}{8mm}
\begin{tabular}{lccr}
\toprule
\textbf{Element} & \textbf{Solar Abundance} & \textbf{LMC Abundance} & \boldmath{$f$}\\
\midrule
C & $2.88 \times 10^{-4}$  & $1.10 \times 10^{-4}$  & $2.62$ \\
O & $4.90 \times 10^{-4}$  & $2.24 \times 10^{-4}$  & $2.19$ \\
N & $6.67 \times 10^{-5}$  & $1.38 \times 10^{-5} $ & $4.90$ \\
S & $1.32 \times 10^{-5}$  & $5.01 \times 10^{-6} $ & $2.63$ \\
\bottomrule
\end{tabular}
\label{tab:elemental abundances ratios}
\end{table}

However, the abundances in Table~\ref{tab:elemental abundances ratios} are the total elemental abundances in both the gas phase and the dust grains. So, in order to obtain the abundances for the gas phase only, we used the initial molecular abundances in Table 1 in~\cite{Nomura2004} to obtain the total galactic gas phase elemental abundances. For instance, the initial abundance of \ce{H2S} is derived from the value in Table~1 of~\cite{Nomura2004} divided by the ratio $f$ for the sulphur element. We arranged the initial abundances so that the total elemental abundances in the gas phase kept the ratio $f$ for each element.
In addition to the molecular species in Table 1 of~\cite{Nomura2004}, we added the initial abundances of $1.00 \times 10^{-8}$ for HNCO, \ce{CH3CN}, \ce{CH3OCH3} and \ce{HCOOCH3}, which are comparable or within the range of the model prediction of grain surface reactions (e.g.,~\cite{Garrod2013}). For HNCO and \ce{CH3CN}, this initial abundance reproduced the observations (see Sections~\ref{subsec3.3} and~\ref{subsec4.5}).

Table~\ref{tab:Ini molec} shows the initial fractional abundances that were used in the model. We assume that the molecular abundances were produced after the evaporation of the grain mantles~\cite{Nomura2004}. These molecular abundances are similar to those of the observations of hot cores, hot corinos and comets, as well as the results of model calculations with grain surface reactions (e.g.,~\cite{Walsh2014,Pauly2018}).

\begin{table}[H]%
\caption{The initial fractional abundances used in the model.}
\setlength{\tabcolsep}{8.25mm}
\begin{tabular}{lccr}
\toprule
\textbf{Species}     & \textbf{Abundance}     & \textbf{Species}  & \textbf{Abundance}  \\
\midrule
 \ce{H+}          &$1.00 \times 10^{-11}$  & \ce{He+}         & $2.50 \times 10^{-12}$ \\
 \ce{H3+}         & $1.00 \times 10^{-9}$  
 & \ce{Fe+}         & $2.40 \times 10^{-8}$\\
 \ce{He}          & $1.00 \times 10^{-1}$  & \ce{S}           & $2.02 \times 10^{-9}$\\
 \ce{C2H2}        & $8.83 \times 10^{-8}$  & \ce{CO2}            & $1.26 \times 10^{-6}$ \\
 \ce{CH4}              & $8.40 \times 10^{-8}$  & \ce{H2CO}         & $8.40 \times 10^{-7}$ \\ 
 \ce{C2H4}             & $8.83 \times 10^{-10}$ & \ce{CH3OH}        & $8.40 \times 10^{-8}$ \\
 \ce{C2H6}            & $8.83 \times 10^{-10}$ & \ce{C2H5OH}    & $8.83 \times 10^{-10}$ \\
 \ce{CO}                & $5.46 \times 10^{-5}$  & \ce{OCS}             & $2.10 \times 10^{-8}$ \\
 \ce{O2}             & $2.07 \times 10^{-7}$  & \ce{N2}             & $1.50 \times 10^{-6}$ \\
 \ce{H2O}           & $1.34 \times 10^{-4}$  & \ce{NH3}          & $1.21 \times 10^{-7}$ \\
 \ce{H2S}          & $3.95 \times 10^{-8}$   &\ce{Si}        & $1.00 \times 10^{-11}$\\
 \ce{HNCO}         & $1.00 \times 10^{-8}$   &\ce{CH3CN}       & $1.00 \times 10^{-8}$\\
\ce{HCOOCH3}      & $1.00 \times 10^{-8}$ &\ce{CH3OCH3}     & $1.00 \times 10^{-8}$\\

 \bottomrule
\end{tabular}
\label{tab:Ini molec}
\end{table}

\section{Results}\label{sec3}

We tested the effects of the variation in the physical parameters on the abundances of the sulphur-bearing species, the nitrogen-bearing species and the COMs in a low-metallicity hot core. We performed model calculations up to the age of $10^{8}$ yrs. The resultant abundances of the molecules were classified into three categories according to their trends with time evolution. The first category is for the parent molecules, such as \ce{OSC}, \ce{HNCO} and \ce{CH3CN}, where their abundances decrease as time increases. The other two categories are for daughter molecules, where one of them contains molecules whose abundances increase with time, for example, \ce{SO}, \ce{SO2}, \ce{NO}, \ce{HCN}, \ce{CN} and \ce{HCO+}, while the other contains molecules whose abundances increase and then decrease, for example, \ce{CS}, \ce{H2CS}, \ce{H2CO} and \ce{HC3N}. For the COMs, such as \ce{CH3OH}, \ce{CH3OCH3} and \ce{HCOOCH3}, they are parent molecules and their abundances decrease with time.
Figures~\ref{fig:1CR}--\ref{fig:3Dn} shows the resulting time evolution of molecular abundances with different physical parameters. We note that in each figure, we fixed two physical parameters because their values reproduce the observation of each molecular abundance the best (see Sections~\ref{subsec3.3} and~\ref{sec4}).



\subsection{The Effect of Cosmic Ray Ionisation Rate Variation}\label{subsec3.1}

\begin{figure}[H]%

\begin{adjustwidth}{-\extralength}{0cm}
\centering 
\minipage{0.43\textwidth}
    \subcaption*{\hspace*{0.2cm}\scriptsize{T = 50 K, n = $2\times10^{5}$ cm$^{-3}$}}
        \vspace*{0.2cm}
    \includegraphics[width=\linewidth]{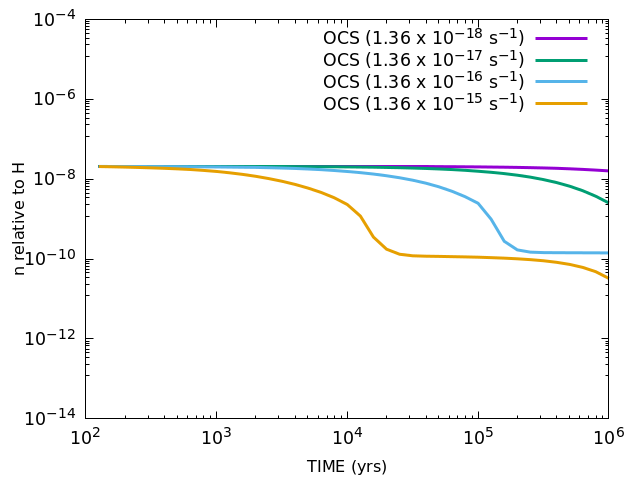}
\endminipage\hfill
\minipage{0.43\textwidth}
    \subcaption*{\hspace*{0.2cm}\scriptsize{T = 50 K, n = $2\times10^{5}$ cm$^{-3}$}}
        \vspace*{0.2cm}
    \includegraphics[width=\linewidth]{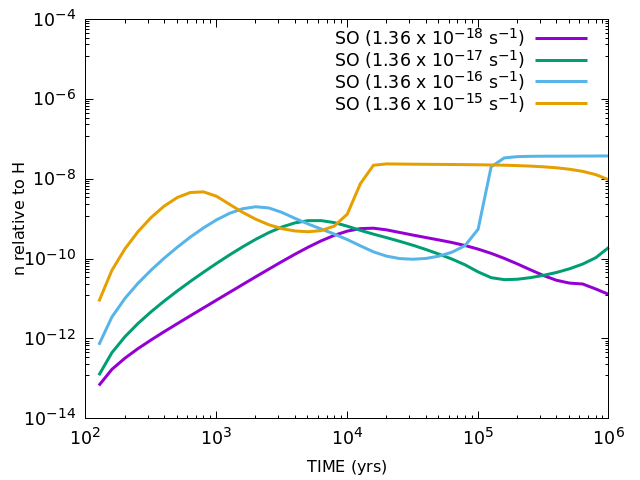}
\endminipage\hfill
\minipage{0.43\textwidth}
    \subcaption*{\hspace*{0.2cm}\scriptsize{T = 50 K, n = $2\times10^{5}$ cm$^{-3}$}}
        \vspace*{0.2cm}
    \includegraphics[width=\linewidth]{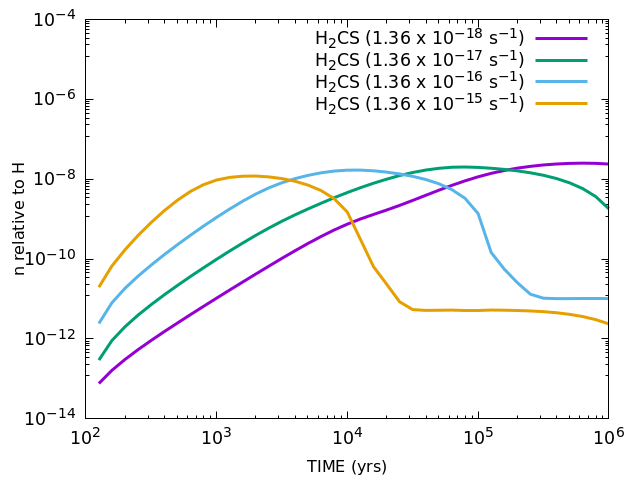}
\endminipage
\\
\minipage{0.43\textwidth}
    \subcaption*{\hspace*{0.2cm}\scriptsize{T = 200 K, n = $2\times10^{7}$ cm$^{-3}$}}
        \vspace*{0.2cm}
    \includegraphics[width=\linewidth]{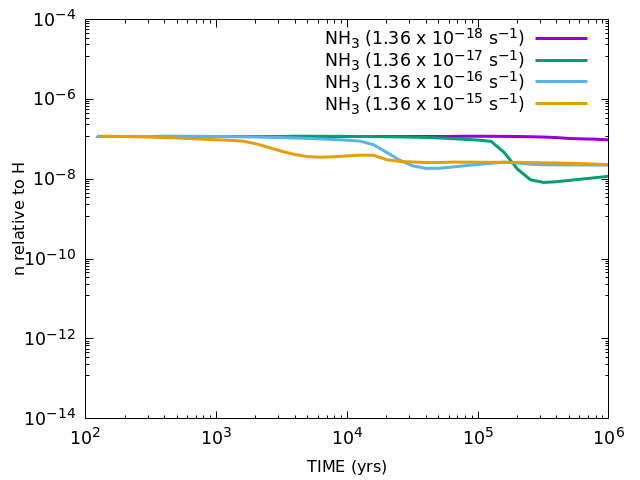}
\endminipage\hfill
\minipage{0.43\textwidth}
    \subcaption*{\hspace*{0.2cm}\scriptsize{T = 200 K, n = $2\times10^{7}$ cm$^{-3}$}}
        \vspace*{0.2cm}
    \includegraphics[width=\linewidth]{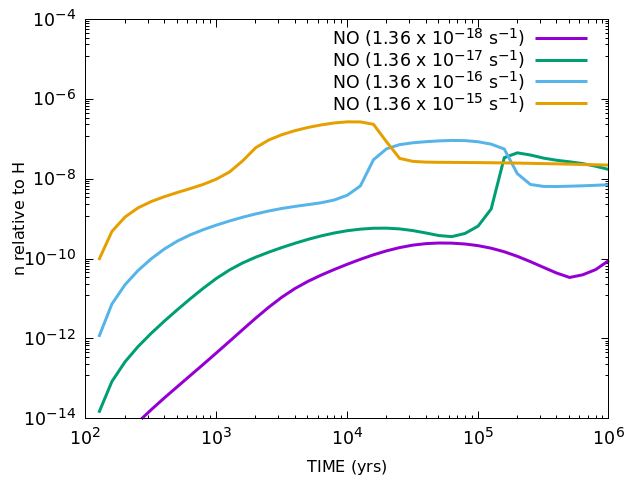}
\endminipage\hfill
\minipage{0.43\textwidth}
    \subcaption*{\hspace*{0.2cm}\scriptsize{T = 200 K, n = $2\times10^{5}$ cm$^{-3}$}}
        \vspace*{0.2cm}
    \includegraphics[width=\linewidth]{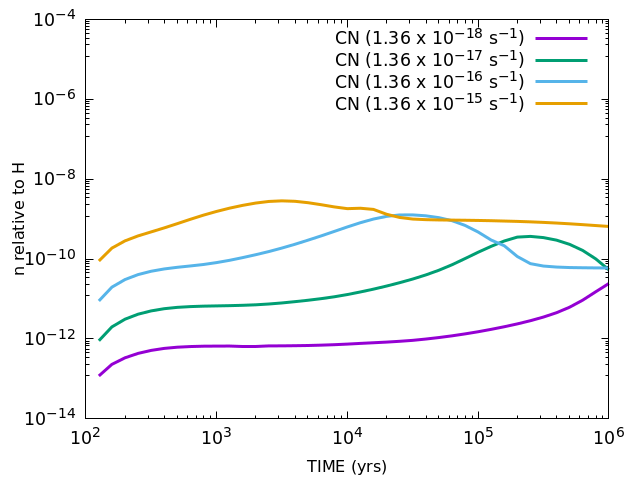}
\endminipage
\\
\minipage{0.43\textwidth}
    \subcaption*{\hspace*{0.2cm}\scriptsize{T = 200 K, n = $2\times10^{7}$ cm$^{-3}$}}
        \vspace*{0.2cm}
    \includegraphics[width=\linewidth]{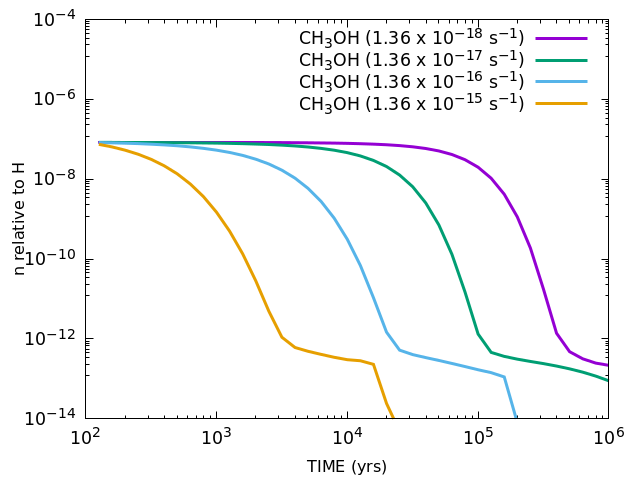}
\endminipage\hfill
\minipage{0.43\textwidth}
    \subcaption*{\hspace*{0.2cm}\scriptsize{T = 200 K, n = $2\times10^{7}$ cm$^{-3}$}}
        \vspace*{0.2cm}
    \includegraphics[width=\linewidth]{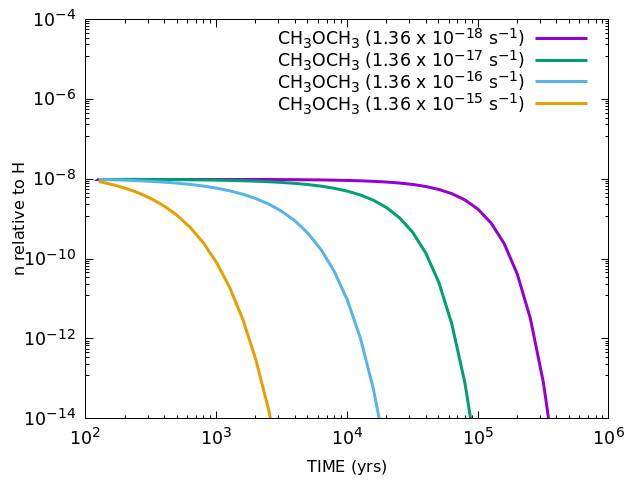}
\endminipage\hfill
\minipage{0.43\textwidth}
    \subcaption*{\hspace*{0.2cm}\scriptsize{T = 200 K, n = $2\times10^{7}$ cm$^{-3}$}}
        \vspace*{0.2cm}
    \includegraphics[width=\linewidth]{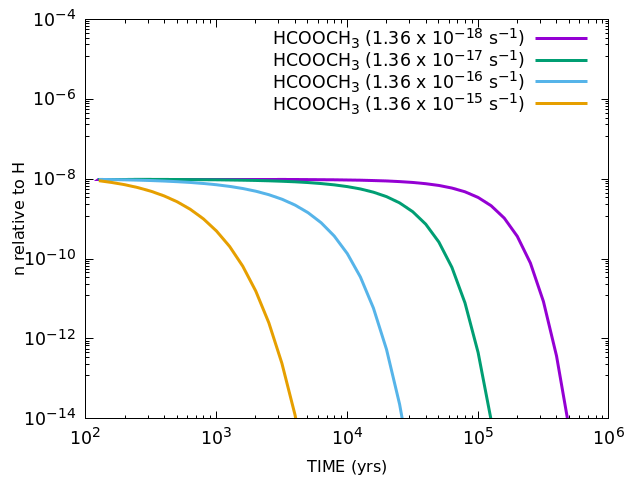}
\endminipage
\end{adjustwidth}
    \caption{{The} effect of varying the CRIR on the time evolution of the molecular abundances of S-bearing species (\textbf{top row}), N-bearing species (\textbf{middle row}) and COMs (\textbf{bottom row}).The lines with different colours show the results for different CRIRs of $1.36 \times 10^{-18}$ s$^{-1}$ (purple), $1.36 \times 10^{-17}$ s$^{-1}$ (green), $1.36 \times 10^{-16}$ s$^{-1}$ (blue) and $1.36 \times 10^{-15}$ s$^{-1}$ (yellow).}
    \label{fig:1CR} 
\end{figure}

The abundances of all species are affected significantly and mainly by CRIR. Figure~\ref{fig:1CR} shows how the abundances of different molecules are affected by varying the CRIRs; as a result, we fixed the values of temperature and density in each graph as they reproduced the observed abundance the best and varied the value of CRIR. Three representative types of classified molecules for each S-bearing species, N-bearing species, and COMs are plotted. It is noticeable that as the value of the CRIR increases, the molecular abundances start to decrease at earlier times, and for some molecules, they reach the steady state at earlier ages as the CRIR increases, except for the COMs.

\subsection{The Effect of Temperature and Density Variation}\label{subsec3.2}

The effect of temperature variation is shown in Figure~\ref{fig:2Temp}, where the values of density and CRIR are fixed. For S-bearing species, the effect of varying the temperature appears at later times for \ce{H2CS} and \ce{OCS}, while it appears at earlier times for \ce{SO}.
For N-bearing species, \ce{NO} and \ce{CN} abundances are affected by the changing temperature at earlier ages. \ce{NH3} abundances show nearly a constant value at different temperatures. 
In  Figure~\ref{fig:3Dn}, we fixed the temperature and the CRIR to test the effect of varying the density on the abundances of different molecules. The results show that the density variation affects the daughter species but hardly affects the parent species, especially at early times. For example, the effect of varying the density is clear at early times for \ce{H2CS} and \ce{SO}. However, the change in density roughly affects \ce{OCS} abundance. Although the effect of density variation appears significantly on the abundances of \ce{NO} and \ce{CN}, this effect is very small on the \ce{NH3} abundance.
Consequently, we found that the parameter that had the strongest impact on the molecular abundances was the CRIR, as confirmed by~\cite{Hatchell1998}.

\begin{figure}[H]%

\begin{adjustwidth}{-\extralength}{0cm}
\centering 
\minipage{0.43\textwidth}
    \subcaption*{\hspace*{0.2cm}\scriptsize{$\zeta = 1.36\times10^{-16}$ s$^{-1}$, n = $2\times10^{5}$ cm$^{-3}$}}
    \vspace*{0.2cm}
    \includegraphics[width=\linewidth]{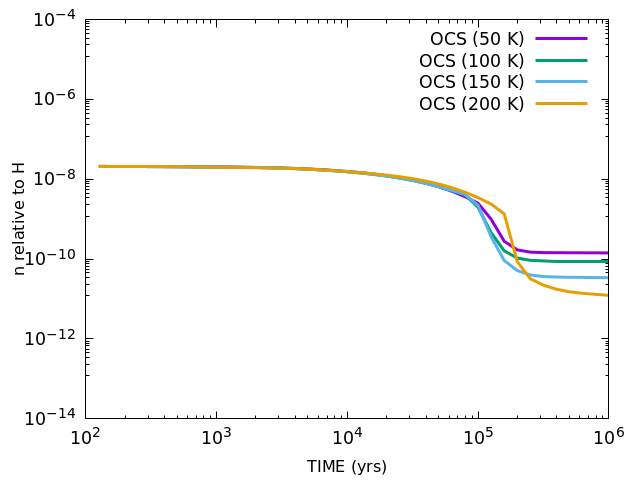}
\endminipage\hfill
\minipage{0.43\textwidth}
    \subcaption*{\hspace*{0.2cm}\scriptsize{$\zeta = 1.36\times10^{-16}$ s$^{-1}$, n = $2\times10^{5}$ cm$^{-3}$}}
    \vspace*{0.2cm}
    \includegraphics[width=\linewidth]{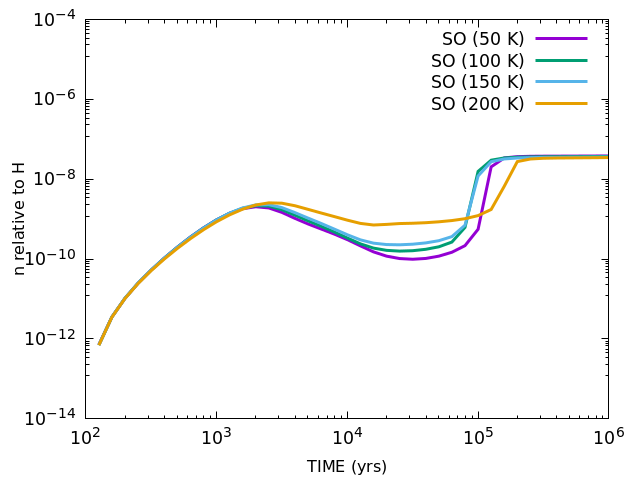}
\endminipage\hfill
\minipage{0.43\textwidth}
    \subcaption*{\hspace*{0.2cm}\scriptsize{$\zeta = 1.36\times10^{-16}$ s$^{-1}$, n = $2\times10^{5}$ cm$^{-3}$}}
    \vspace*{0.2cm}
    \includegraphics[width=\linewidth]{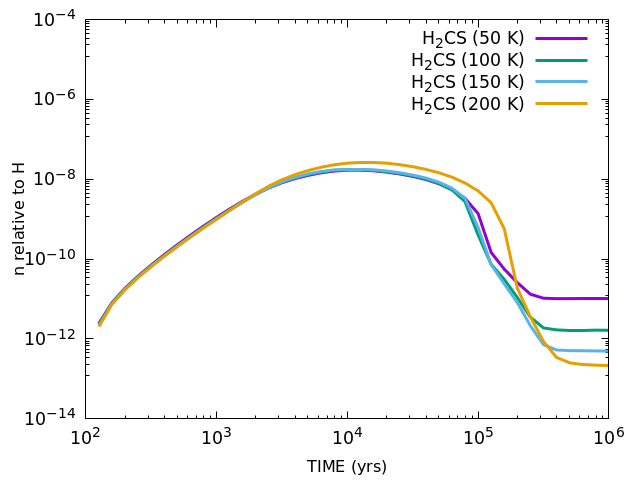}
\endminipage
\end{adjustwidth}

    \label{fig:2Temp}
\end{figure}

\vspace{-12pt}

\begin{figure}[H]\ContinuedFloat

\begin{adjustwidth}{-\extralength}{0cm}
\centering 
\minipage{0.43\textwidth}
    \subcaption*{\hspace*{0.2cm}\scriptsize{$\zeta = 1.36\times10^{-16}$ s$^{-1}$, n = $2\times10^{7}$ cm$^{-3}$}}
    \vspace*{0.2cm}
    \includegraphics[width=\linewidth]{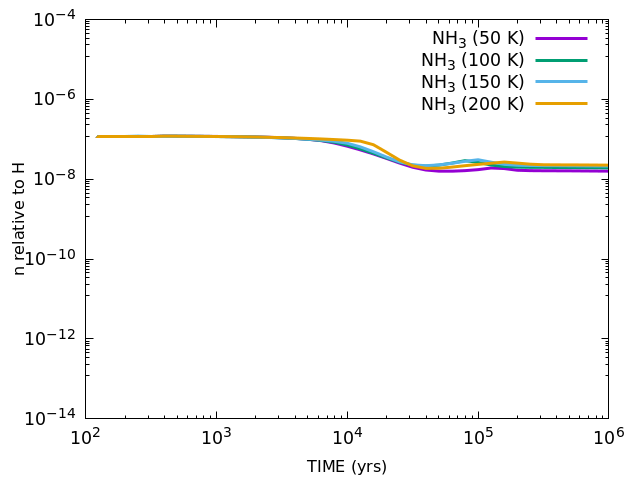}
\endminipage\hfill
\minipage{0.43\textwidth}
    \subcaption*{\hspace*{0.2cm}\scriptsize{$\zeta = 1.36\times10^{-16}$ s$^{-1}$, n = $2\times10^{7}$ cm$^{-3}$}}
    \vspace*{0.2cm}
    \includegraphics[width=\linewidth]{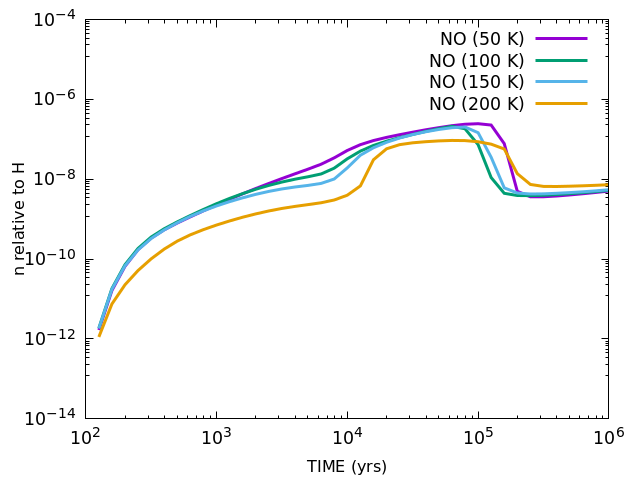}
\endminipage\hfill
\minipage{0.43\textwidth}
    \subcaption*{\hspace*{0.2cm}\scriptsize{$\zeta = 1.36\times10^{-16}$ s$^{-1}$, n = $2\times10^{5}$ cm$^{-3}$}}
    \vspace*{0.2cm}
    \includegraphics[width=\linewidth]{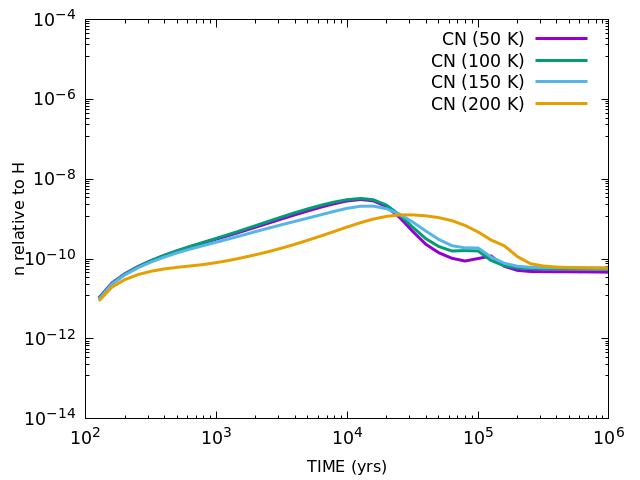}
\endminipage
\\
\minipage{0.43\textwidth}
    \subcaption*{\hspace*{0.2cm}\scriptsize{$\zeta = 1.36\times10^{-16}$ s$^{-1}$, n = $2\times10^{7}$ cm$^{-3}$}}
    \vspace*{0.2cm}
    \includegraphics[width=\linewidth]{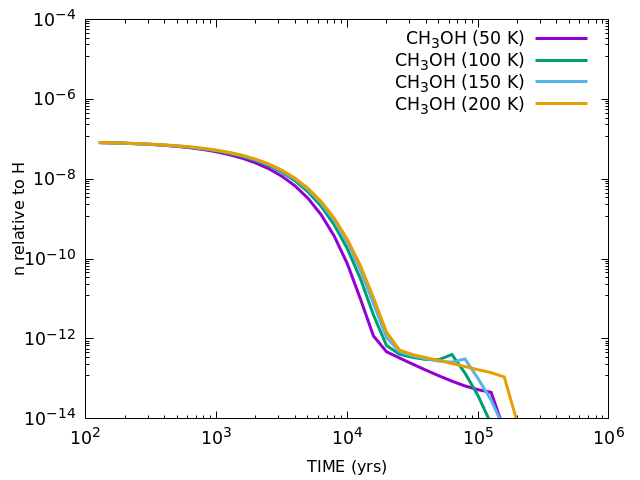}
\endminipage\hfill
\minipage{0.43\textwidth}
    \subcaption*{\hspace*{0.2cm}\scriptsize{$\zeta = 1.36\times10^{-16}$ s$^{-1}$, n = $2\times10^{7}$ cm$^{-3}$}}
    \vspace*{0.2cm}
    \includegraphics[width=\linewidth]{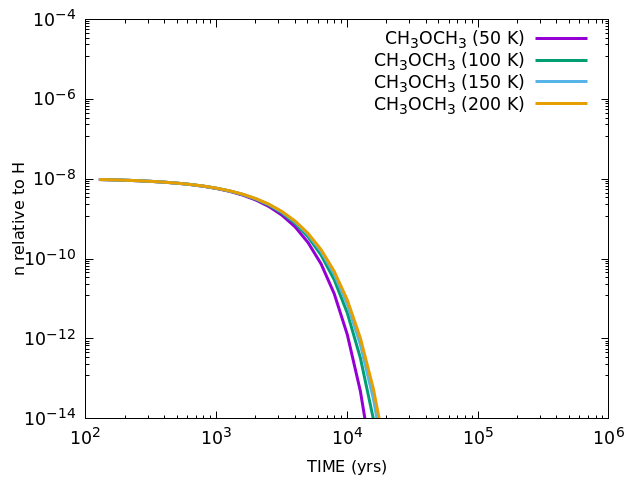}
\endminipage\hfill
\minipage{0.43\textwidth}
    \subcaption*{\hspace*{0.2cm}\scriptsize{$\zeta = 1.36\times10^{-16}$ s$^{-1}$, n = $2\times10^{7}$ cm$^{-3}$}}
    \vspace*{0.2cm}
    \includegraphics[width=\linewidth]{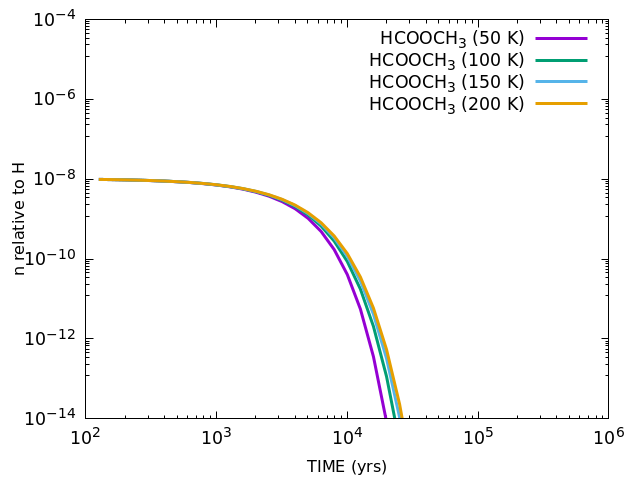}
\endminipage
\end{adjustwidth}
    \caption{{The} same as Figure~\ref{fig:1CR} but for different temperatures. The lines with different colours show the results for different temperatures of 50 K (purple), 100 K (green), 150 K (blue) and 200 K (yellow).}
    \label{fig:2Temp}   
\end{figure}

\begin{figure}[H]%

\begin{adjustwidth}{-\extralength}{0cm}
\centering 
\minipage{0.43\textwidth}
    \subcaption*{\scriptsize{T = 50 K, $\zeta = 1.36\times10^{-16}$ s$^{-1}$}}
    \vspace*{0.1cm}
    \includegraphics[width=\linewidth]{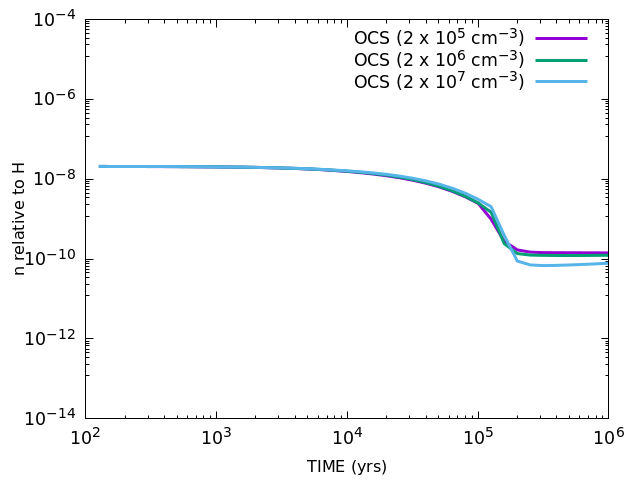}
\endminipage\hfill
\minipage{0.43\textwidth}
    \subcaption*{\scriptsize{T = 50 K, $\zeta = 1.36\times10^{-16}$ s$^{-1}$}}
    \vspace*{0.1cm}
    \includegraphics[width=\linewidth]{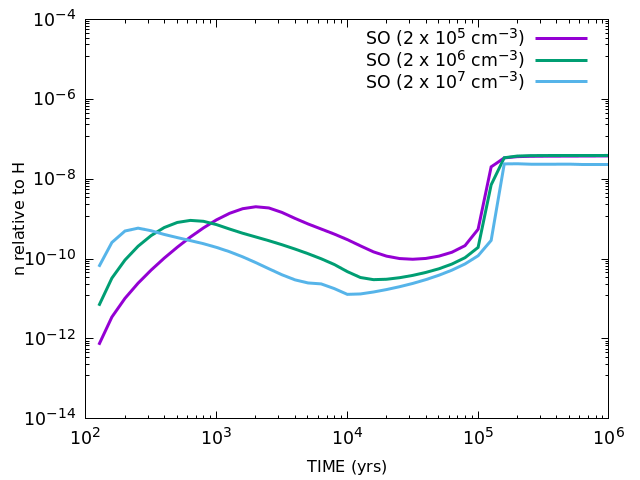}
\endminipage\hfill
\minipage{0.43\textwidth}
    \subcaption*{\scriptsize{T = 50 K, $\zeta = 1.36\times10^{-16}$ s$^{-1}$}}
    \vspace*{0.1cm}
    \includegraphics[width=\linewidth]{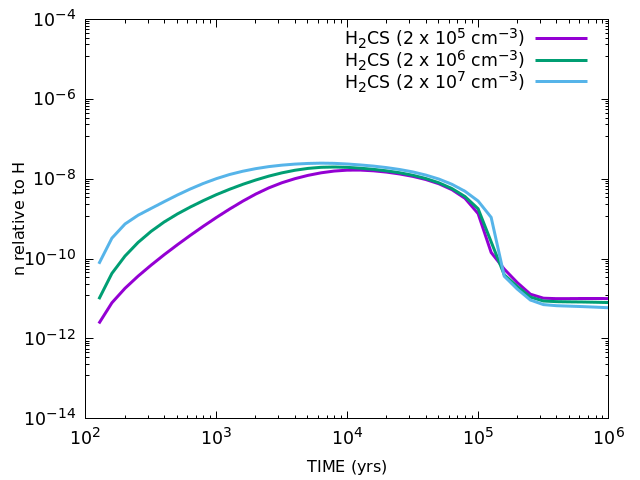}
\endminipage
\\
\minipage{0.43\textwidth}
    \subcaption*{\scriptsize{T = 200 K, $\zeta = 1.36\times10^{-16}$ s$^{-1}$}}
    \vspace*{0.1cm}
    \includegraphics[width=\linewidth]{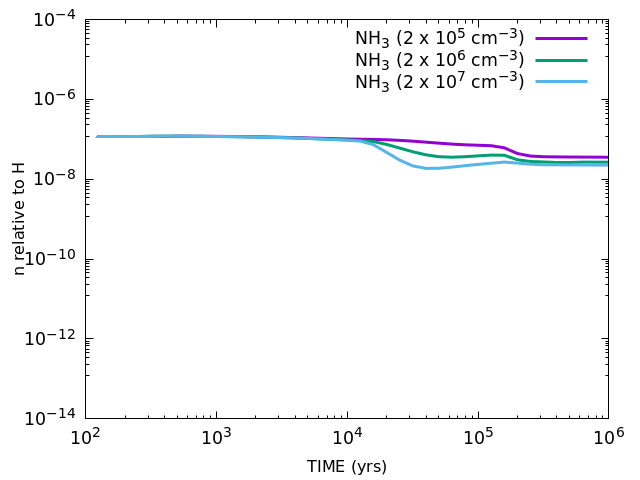}
\endminipage\hfill
\minipage{0.43\textwidth}
    \subcaption*{\scriptsize{T = 200 K, $\zeta = 1.36\times10^{-16}$ s$^{-1}$}}
    \vspace*{0.1cm}
    \includegraphics[width=\linewidth]{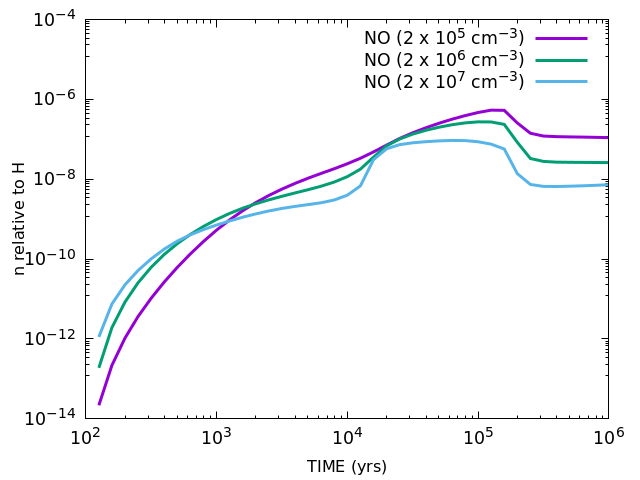}
\endminipage\hfill
\minipage{0.43\textwidth}
    \subcaption*{\scriptsize{T = 200 K, $\zeta = 1.36\times10^{-16}$ s$^{-1}$}}
    \vspace*{0.1cm}
    \includegraphics[width=\linewidth]{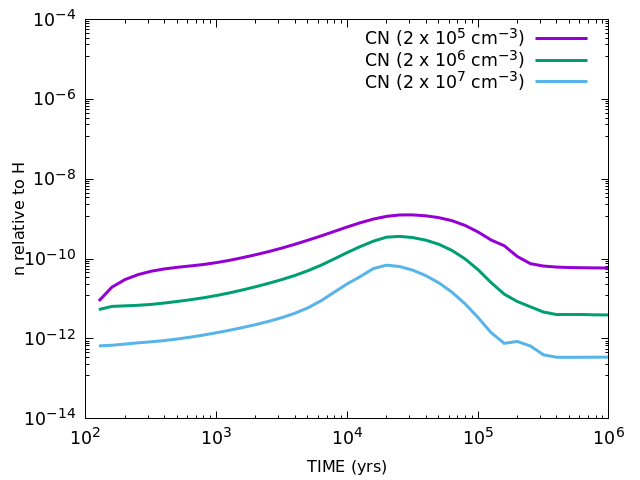}
\endminipage
\end{adjustwidth}
  \label{fig:3Dn}
\end{figure}

\vspace{-12pt}

\begin{figure}[H]\ContinuedFloat
\begin{adjustwidth}{-\extralength}{0cm}
\centering 

\minipage{0.43\textwidth}
    \subcaption*{\scriptsize{T = 200 K, $\zeta = 1.36\times10^{-16}$ s$^{-1}$}}
    \vspace*{0.1cm}
    \includegraphics[width=\linewidth]{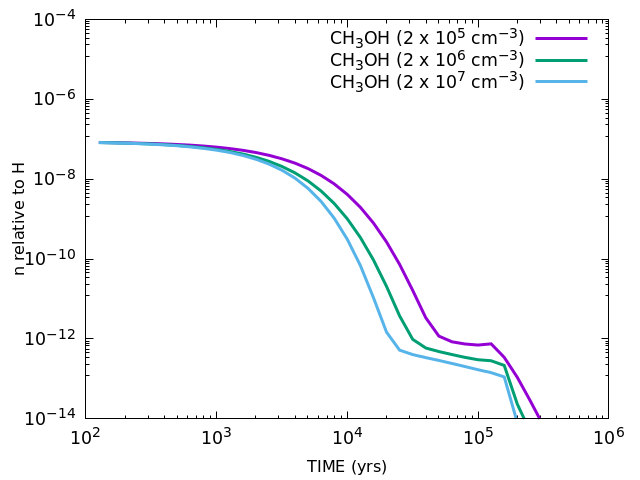}
\endminipage\hfill
\minipage{0.43\textwidth}
    \subcaption*{\scriptsize{T = 200 K, $\zeta = 1.36\times10^{-16}$ s$^{-1}$}}
    \vspace*{0.1cm}
    \includegraphics[width=\linewidth]{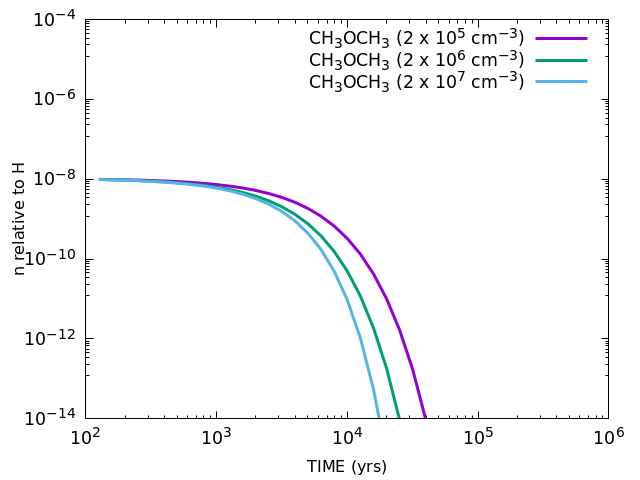}
\endminipage\hfill
\minipage{0.43\textwidth}
    \subcaption*{\scriptsize{T = 200 K, $\zeta = 1.36\times10^{-16}$ s$^{-1}$}}
    \vspace*{0.1cm}
    \includegraphics[width=\linewidth]{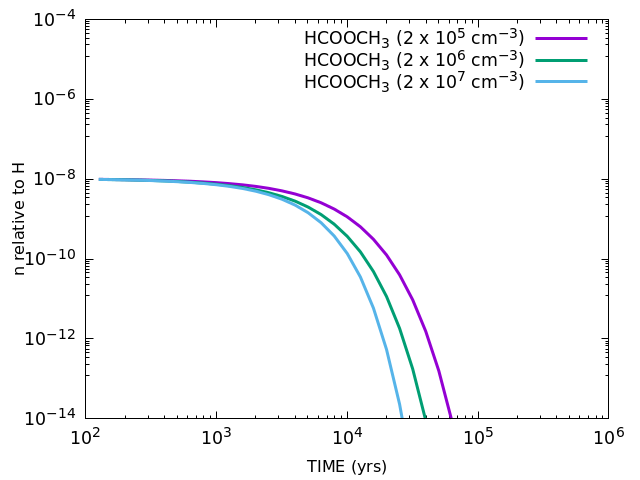}
\endminipage
\end{adjustwidth}
    \caption{{The} same as Figure~\ref{fig:1CR} but for different densities. The lines with different colors show the results for different densities of $2 \times 10^5$ cm$^{-3}$ (purple), $2 \times 10^6$ cm$^{-3}$ (green) and $2 \times 10^7$ cm$^{-3}$ (blue).}   
    \label{fig:3Dn}
\end{figure}

\subsection{Comparing Results with Observations}\label{subsec3.3}
The resultant abundances of these molecules were compared with their observational abundances in  Figure~\ref{fig:4Compare}. The graphs on the left-hand side of Figure~\ref{fig:4Compare} contain the observational abundances of different molecules from LMC hot cores together with the abundances obtained by model calculations using $\zeta = 1.36 \times 10^{-16}$ s$^{-1}$, while the plots on the right-hand side represent the deviation in the molecular abundances between the observations and the model calculations at different times for various CRIRs. It is a similar way to that used in~\cite{Entekhabi2022}. This deviation is given as follows:
\begin{equation*}
   \chi^{2} (t,\zeta) = \sum\limits_{i}(\log_{10}[X_i]_{\rm obs\ } - \log_{10}[X_{i}(t,\zeta)]_{\rm theory\ })^{2}, 
\end{equation*}
where $[X_i]_{\rm obs\ }$ represent the observed abundances, and $[X_{i}(t,\zeta)]_{\rm theory\ }$ represents the model abundances.
Sets of sulphur-bearing and nitrogen-bearing species and COMs whose abundances were measured by observations were treated together, and $\chi^{2}(t,\zeta)$ were calculated by summing up the values for three molecules in each figure. $\chi^{2}(t,\zeta)$ was plotted for the molecules in Table~\ref{tab:obs mol} to determine the time and the CRIR at which the theoretical abundances fit the best with the observations. Table~\ref{tab:obs mol} contains the observed abundances of the S-bearing species, the N-bearing species and the COMs that were used in the comparison. We note that in each figure, we fixed temperature and density as their values reproduced the observations of a set of molecular abundances well (see Section~\ref{sec4}). These figures show that the abundances of daughter species deviated from the observed values in early times, while those of parent species and daughter species whose abundances decreased with time deviated from the observed values at a later time. Therefore, the results of the model calculations fit the observed values at around 10$^5$ yrs and $\zeta = 1.36 \times 10^{-16}$ s$^{-1}$, except for the COMs. Even if we excluded HNCO and CH3CN, this result is not affected. We note that the model calculations fit the observed values with lower CRIRs and later times as well, but it is unlikely that the ages of hot cores are older than $10^5$ yrs (e.g.,~\cite{York2004}; see Section~\ref{subsec4.2}). $\chi^2$ values for COMs are very large, which is suggestive of the large deviation in the resulted abundances from the observed ones, as shown in Figure~\ref{fig:4Compare}. The deviation between the model and the observations of COMs are discussed in Section~\ref{subsec4.5}.
\begin{table}[H]%
\caption{The observed abundances used in the comparison.}
\label{tab:obs mol}
\setlength{\tabcolsep}{12.6mm}
\begin{tabular}{lccr}
\toprule
\textbf{Species}     & \textbf{Abundance}     & \textbf{LMC Hot Core}  \\
\midrule
 \ce{CS}             & $3.33 \times 10^{-10}$  &                  \\
 \ce{OCS}            & $1.20 \times 10^{-9}$  &      
  \\
 \ce{SO2}            & $2.00 \times 10^{-8}$  &                  \\
 \ce{NO}             & $7.30 \times 10^{-9}$  &                  \\
 \ce{HCN}            & $7.70 \times 10^{-10}$  &      ST16       \\  
 \ce{CN}             & $8.00 \times 10^{-11}$ &                    \\
 \ce{HNCO}           & $6.60 \times 10^{-11}$ &                   \\
 \ce{HC3N}           & $1.50 \times 10^{-11}$  &                  \\
 \ce{CH3CN}          & $4.10 \times 10^{-11}$  &                  \\
 \ce{CH3OH}          & $6.66 \times 10^{-9}$  &                  \\
 \midrule
 \ce{CH3OCH3}        & $2.10 \times 10^{-9}$   &      N105-2A     \\
 \ce{HCOOCH3}        & $4.00 \times 10^{-9}$   &      N113 A1 \\
 \bottomrule
\end{tabular}

	\noindent{\footnotesize{Note: The observation data are from~\cite{Shimon2020,Sewilo2022,Sewilo2018} for ST16, N105-2A, N113 A1, respectively.}}
\end{table}

\vspace{-18pt}

\begin{figure}[H]%
\minipage{0.45\textwidth}
    \subcaption*{\hspace*{0.5cm} T = 50 K, n = $2\times10^{5}$ cm$^{-3}$}
    \vspace*{0.1cm}
    \includegraphics[width=\linewidth]{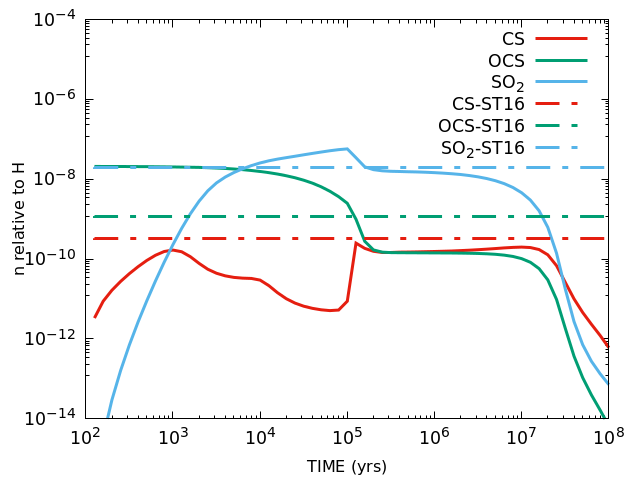}
\endminipage\hfill
\minipage{0.5\textwidth}
    \subcaption*{\hspace*{0.7cm} T = 50 K, n = $2\times10^{5}$ cm$^{-3}$}
    \vspace*{0.1cm}
    \includegraphics[width=\linewidth]{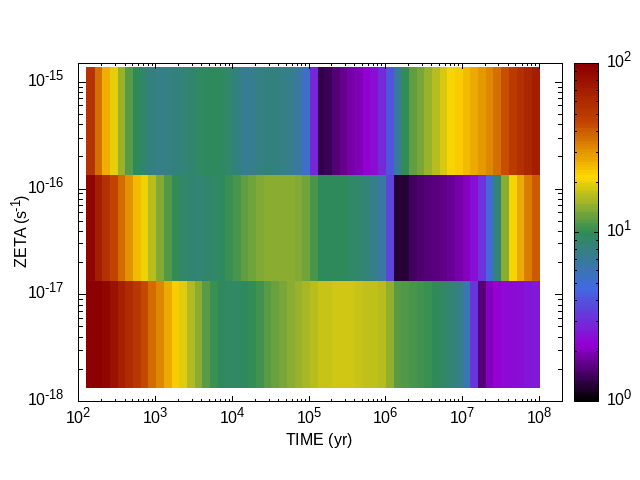}
    \label{fig:my_label}
\endminipage

\vspace{-18pt}
    \label{fig:4Compare}
\end{figure}

\begin{figure}[H]\ContinuedFloat
\minipage{0.45\textwidth}
    \subcaption*{\hspace*{0.5cm} T = 200 K, n = $2\times10^{7}$ cm$^{-3}$}
    \vspace*{0.1cm}
    \includegraphics[width=\linewidth]{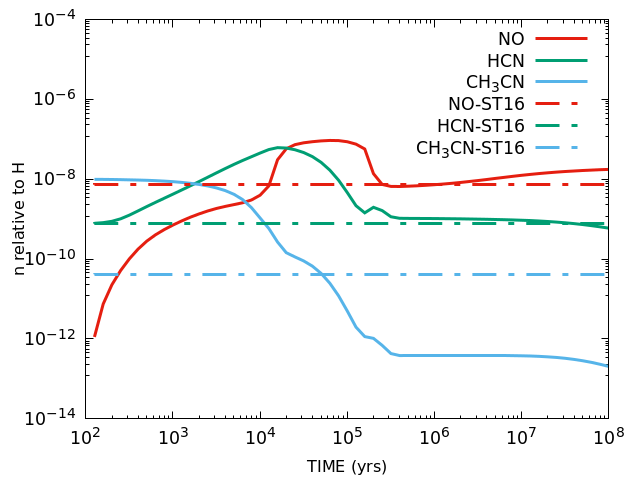}
    \label{fig:my_label}
\endminipage\hfill
\minipage{0.5\textwidth}
    \subcaption*{\hspace*{0.7cm} T = 200 K, n = $2\times10^{7}$ cm$^{-3}$}
    \vspace*{0.1cm}
    \includegraphics[width=\linewidth]{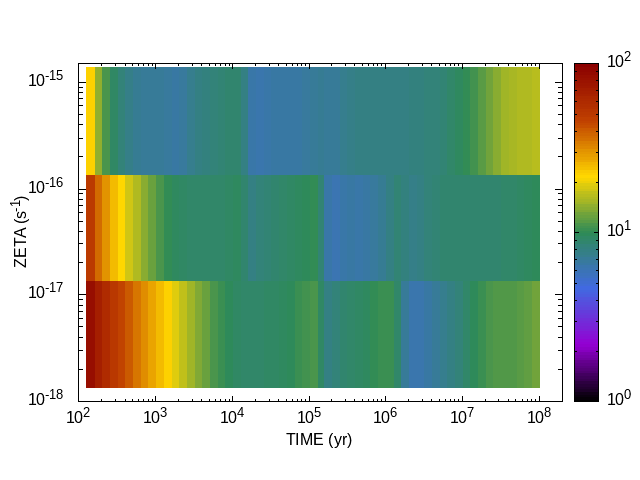}
\endminipage
\\
\minipage{0.45\textwidth}
    \subcaption*{\hspace*{0.5cm} T = 100 K, n = $2\times10^{5}$ cm$^{-3}$}
    \vspace*{0.1cm}
    \includegraphics[width=\linewidth]{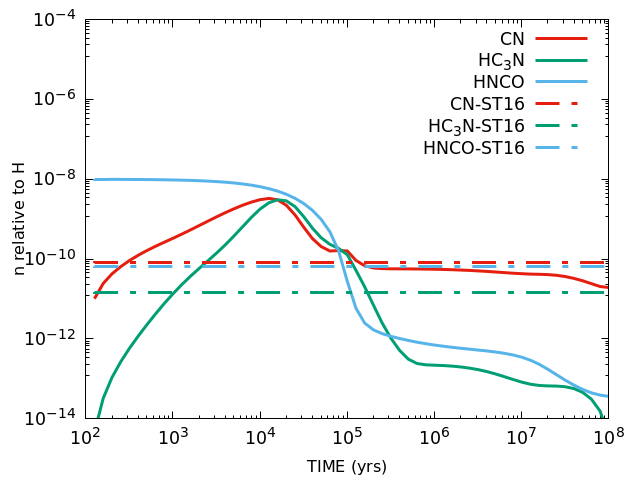}
\endminipage\hfill
\minipage{0.5\textwidth}
    \subcaption*{\hspace*{0.7cm} T = 100 K, n = $2\times10^{5}$ cm$^{-3}$}
    \vspace*{0.1cm}
    \includegraphics[width=\linewidth]{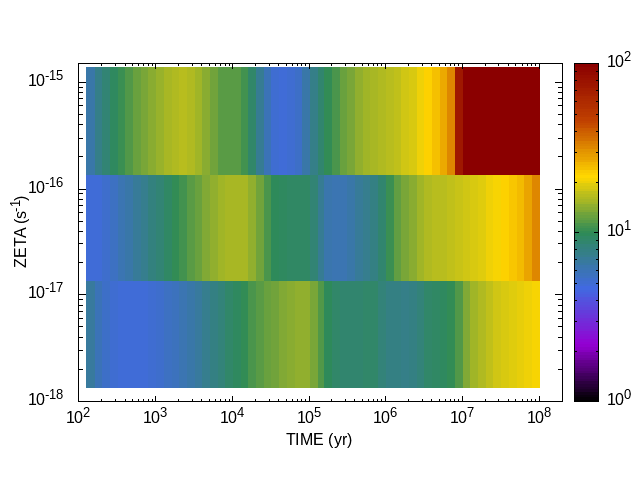}    
\endminipage
\\
\minipage{0.45\textwidth}
    \subcaption*{\hspace*{0.5cm} T = 200 K, n = $2\times10^{7}$ cm$^{-3}$}
    \vspace*{0.1cm}
    \includegraphics[width=\linewidth]{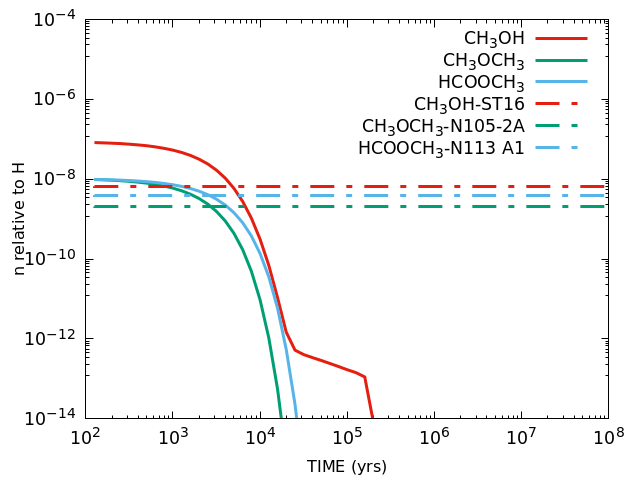}    
    \label{fig:my_label}
\endminipage\hfill
\minipage{0.5\textwidth}
    \subcaption*{\hspace*{0.7cm} T = 200 K, n = $2\times10^{7}$ cm$^{-3}$}
    \vspace*{0.1cm}
    \includegraphics[width=\linewidth]{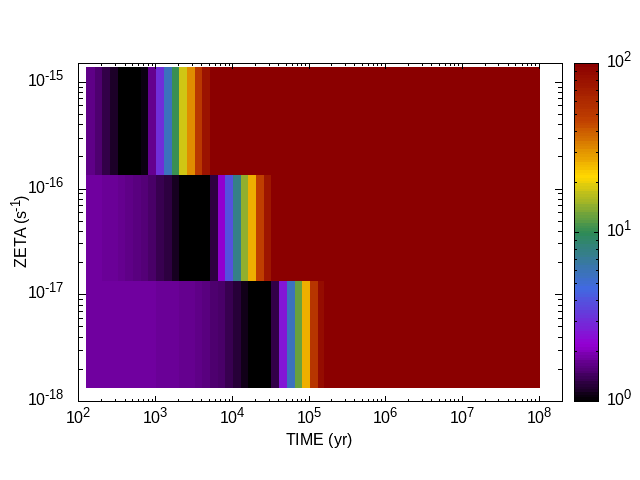}
    \label{fig:my_label}
\endminipage
\vspace{-18pt}
    \caption{({\textbf{Left}}) {The} results of model calculations for the time evolution (solid lines) of the abundances of S-bearing species (first row), N-bearing species (second and third rows) and COMs (fourth row) using \mbox{$\zeta=1.36\times 10^{-16}$ s$^{-1}$}, together with the observed molecular abundances (dot-dashed lines). ({\textbf{Right}}) The deviation in molecular abundances between the observations and the model calculations at different times for different CRIRs $\chi^{2}(t, \zeta)$.} 

    \label{fig:4Compare}
\end{figure}

\section{Discussion}\label{sec4}
In general, our model fits well with the observations from \citet{Shimon2020} and \citet{Sewilo2022}. 
In this section, we discuss in detail the magnitude of CRIR that made our resulted abundances fit with the observations, the timescales of the hot cores in addition to the abundances of S-bearing species, the N-bearing species and the COMs that resulted from our model.

\subsection{The Cosmic-Ray Ionisation Rate}\label{subsec4.1}

Our resultant molecular abundances show that when the CRIR value equals \linebreak$1.36 \times 10^{-16}$ s$^{-1}$, the resultant abundances fit with the observations the best. Previous works used different values of the CRIRs. {Refs.}~\cite{Acharyya2015,Pauly2018} used the value of \mbox{$1.3 \times 10^{-17}$ s$^{-1}$}; the former modelled the chemistry in cold, dense regions in the LMC, while the latter modelled the chemistry in the envelopes of massive young stellar objects. However, the regional variability of CRIRs is significant as is mentioned in~\cite{Pauly2018}.  

Our value of CRIR is high in the LMC hot core, and this may be due to the presence of an SNR in the neighbourhood of the hot core, where about 60 SNRs were observed in the LMC (e.g.,~\cite{Vaupr2014, Maggi2016, Ou2018}). The hot core, ST16, is located at (RA, DEC) = (05:19:12.31, $-$69:9:7.3). According to the list in~\cite{Maggi2016}, some SNRs are located near ST16, for example, J0519-6926, J0519-6902 and J0513-6912. They do not overlap with ST16, but cosmic-ray acceleration at the SNRs may affect the CRIR at ST16. Another possible reason for this high CRIR is because the CRIRs inside the star-forming region; for example,~\mbox{\cite{Morales2014,Cabedo2023}}. {Ref.}~\cite{Morales2014} deduced that the CRIR in the star-forming region NGC 6334 I is an order of magnitude higher than the standard galactic value using the abundance of the molecular ions, \ce{HCO+} and \ce{N2H+}. In the case of~\cite{Cabedo2023}, they found that the CRIR magnitude could reach $10^{-14}$ s$^{-1}$ at a distance of a few hundred au from the central protostar in the Bok Globule B335. They used the deuteration fraction of the gas and its ionisation fraction to obtain the CRIR. Our results may suggest that some cosmic-ray acceleration mechanisms are also efficient in LMC hot cores.

\subsection{Timescale of Hot Cores}\label{subsec4.2}

Our results show that the model calculations of molecular abundance fit
the observations well when the age of the hot cores is $\sim$$10^5$ yrs. This result is consistent with our model's prediction for hot cores in the Milky Way (e.g.,~\cite{Herbst2009}). A comparison between the observed molecular abundances and model calculations suggests that the timescales of the Orion Hot Cores and the Compact Ridge are $10^4$--$10^5$ yrs (e.g.,~\cite{Charnley1992}). 
The determination of the timescales of hot cores was also performed by~\cite{Wilner2001}. They traced the hot cores by \ce{CH3CN} emissions in the star-forming region W49N; then, they compared the ratio between the number of observed hot cores and the number of ultracompact HII regions. They found that the hot core lifetime is comparable to the lifetime of the ultracompact HII regions, that is, about $10^5$ yrs.
\cite{Hatchell1998} found that the timescales of hot cores varied with the temperature and the CRIR. This variation can be investigated by noticing the time at which \ce{H2S} abundance starts to fall and the time at which \ce{SO2} abundance reaches its peak. {Ref.}~\cite{Hatchell1998} noticed that as the temperature and CRIR increase, both \ce{H2S} destruction and \ce{SO2} formation start at earlier times. Therefore, the age of hot cores can range from a few thousand to around 10$^5$ yrs.

\subsection{Sulphur-Bearing Species}\label{subsec4.3}
The abundances of sulphur-bearing species, such as \ce{SO}, \ce{SO2}, \ce{CS}, \ce{H2CS} and \ce{OCS}, change significantly within 10$^5$ yrs after the evaporation of ices when the ionisation rate is 1.36 $\times 10^{-16}$ s$^{-1}$. Our abundances obtained for the S-bearing species are comparable with the abundances observed from the LMC hot cores ST16, ST11, N113 A1 and N105-2 A. As described in~\cite{Charnley1997, Nomura2004}, sulphur-bearing species originate from \ce{H2S} since this was the most abundant sulphur-species at the initial condition. At relatively lower temperatures, for example, T = 50 K, the main destruction process of \ce{H2S} is protonation by \ce{H3O+}. The recombination of \ce{H3S+} produces SH and S, and the produced S reacts with OH to produce SO and \ce{SO2}. 
At a chemical equilibrium state, SO can be formed from the reaction of \ce{SO2} with \ce{C}. However, the sulphur element is originally from \ce{H2S}.


The resultant \ce{OCS} abundance from modelling at the age of $10^5$ yrs indicates that \ce{OCS} has low dependency on temperature. Both \ce{CS} and \ce{H2CS} fit better at lower densities and temperatures. They are formed from the reaction of \ce{H3CS+} with an electron, and in addition, \ce{H2CS} can be destroyed by CR photoionisation to produce \ce{CS}.

{Ref.}~\cite{Shimon2020} adopted the rotational temperature of 50~K for SO, \ce{SO2}, CS and OCS, based on the observations. When comparing the observational abundances of these molecules with our results, we found that SO, \ce{SO2} and OCS fits with observations at 50~K (as well as 100~K, 150~K and 200~K), which is consistent with the temperature adopted in~\cite{Shimon2020}. The resultant CS abundance also fit with observations at lower temperatures (50~K and 100~K) and low densities. Meanwhile,~\cite{Shimon2020} adopted the rotational temperature of 100~K for \ce{H2CS}. The resulting abundance agreed well with observations when the temperature was relatively low (50 and 100~K), but it deviated significantly at higher temperatures, which is also consistent with the temperature adopted in~\cite{Shimon2020}. When the temperature increased, the trend of the resulted \ce{H2CS} deviated significantly from the observations.  
 
\subsection{Nitrogen-Bearing Species}\label{subsec4.4}

\ce{NO} is believed to be formed in the inner regions of the hot cores~\cite{Shimon2020}, and our results for NO abundance agree with this since \ce{NO} fits best at the highest density (at the order of $10^7$ cm$^{-3}$) and at the highest temperature (200 K). \ce{HNO} was found to be the main reservoir for \ce{NO} at this core age ($10^5$ yrs); where \ce{HNO} reacts with either H or O, in both reactions, it results in \ce{NO}. Unlike \ce{NO}, \ce{CN} is formed in the outer regions that have the lowest density, and it is produced from the reaction of \ce{OCN} with H.

Since the abundances of N-bearing species increases as the temperature increases~\cite{Rodgers2001}, we found that \ce{HCN} and \ce{HC3N} abundances resulted from the model fit with observational abundances when the temperature was below 150 K. Their abundances increased above the observational abundances for the temperatures that were equal to or higher than 150 K.

\begin{figure}[H]%
    
\minipage{0.45\textwidth}
    \subcaption*{\footnotesize{T = 200 K, n = $2\times10^{7}$ cm$^{-3}$, $\zeta = 1.36\times10^{-16}$ s$^{-1}$}}
    \vspace*{0.1cm}
    \includegraphics[width=\linewidth]{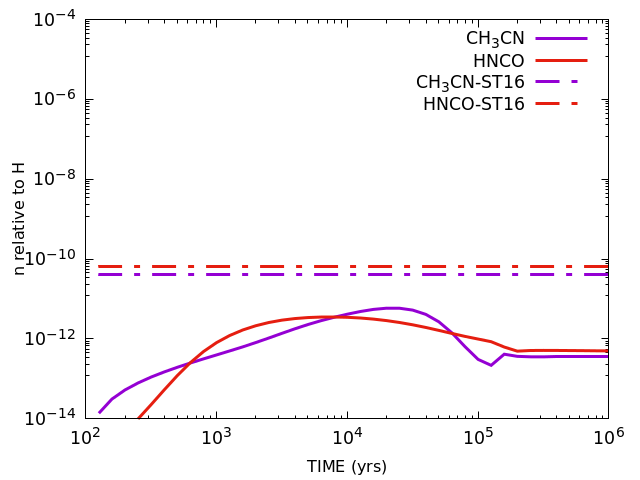}    
    \label{fig:my_label}
\endminipage\hfill
\minipage{0.45\textwidth}
    \subcaption*{\footnotesize{T = 200 K, n = $2\times10^{7}$ cm$^{-3}$, $\zeta = 1.36\times10^{-16}$ s$^{-1}$}}
    \vspace*{0.1cm}
    \includegraphics[width=\linewidth]{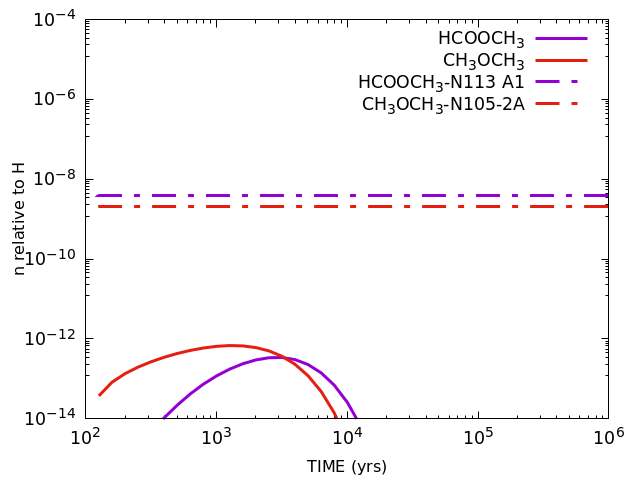}
    \label{fig:my_label}
\endminipage
    \caption{{The} results of model calculations of the time evolution (solid lines) together with the observed molecular abundances (dot-dashed lines), when the initial abundances of N-bearing COMs ({\textbf{left}}) and O-bearing COMs ({\textbf{right}}) were zero.}  
    \label{fig:5settozero}

    \end{figure}

\ce{HCN} is produced from the dissociative recombination reaction of \ce{HCNH+} with \ce{e-}, while \ce{HC3N} is formed from the reaction of \ce{H2C3N+} with an electron (as mentioned in~\cite{Nomura2004} and from the reaction of \ce{CH2CCH} with nitrogen. The abundance of \ce{CH3CN} depends mainly on the abundance of \ce{HCN} ~\cite{Nomura2004}, and it is formed from the reaction of \ce{CH3CNH+} with an electron. In the gas phase, \ce{HNCO} is formed mainly by the reaction of \ce{HNCOH+} and \ce{H2NCO+}, each with an electron.
If we did not input HNCO and \ce{CH3CN} initially, the model calculations did not reproduce the observational abundance, as shown in {Figure}~\ref{fig:5settozero}. Thus, we needed the initial HNCO and \ce{CH3CN} abundances of 1.0$\times 10^{-8}$ to fit with the observations, suggesting that both HNCO and \ce{CH3CN} are formed by grain surface reactions.  

{Ref.}~\cite{Shimon2020} adopted the rotational temperatures of 100 K for N-bearing species, such as NO, HNCO and \ce{HC3N}. When examining the resultant abundances of the N-bearing species, it was found that density had a greater effect on their abundances rather than temperature. Therefore, the calculated results of both NO and HNCO abundances fit the observations at nearly all temperatures, which did not conflict with the temperature adopted in~\cite{Shimon2020}. The abundance of CN also agrees with our observations at all temperatures, which does not conflict with the adopted rotational temperature of 50 K for CN in~\cite{Shimon2020}.

\begin{figure}[H]%
    
\minipage{0.45\textwidth}
    \subcaption*{\footnotesize{T = 200 K, n = $2\times10^{7}$ cm$^{-3}$}}
    \vspace*{0.1cm}
    \includegraphics[width=\linewidth]{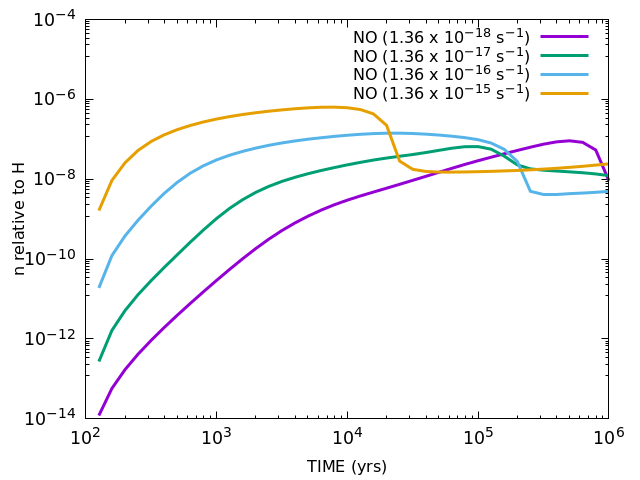}    
    \label{fig:my_label}
\endminipage\hfill
\minipage{0.45\textwidth}
    \subcaption*{\footnotesize{T = 200 K, n = $2\times10^{5}$ cm$^{-3}$}}
    \vspace*{0.1cm}
    \includegraphics[width=\linewidth]{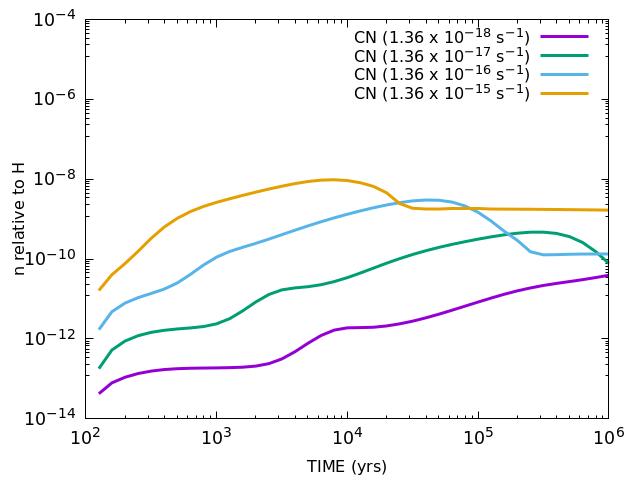}
    \label{fig:my_label}
\endminipage
    \caption{{The} time evolution of NO and CN abundances with the same physical parameters as Figure~\ref{fig:1CR} but with the initial abundances for a galactic hot core.}
    \label{fig:6MW-NO-CN} 

    \end{figure}

{We note that when we compared the time evolution of molecular abundances between the Milky Way and the LMC elemental abundances, most of the molecules showed similar trends, but some of them showed the non-linear effects of chemical reactions. NO is one of the typical examples. Figure~\ref{fig:6MW-NO-CN} shows the time evolution of NO and CN abundances, where we used the same physical parameters as in Figure~\ref{fig:1CR}, but with the initial abundances for a galactic hot core, as shown in Table 1 of~\cite{Nomura2004}. The figure shows that the time evolution of CN abundance behaved similarly while that of NO abundance behaved differently based on the Milky Way and the LMC elemental abundances. Observations at different environments, especially with different CRIRs both in the Milky Way and the LMC, could help understand the formation pathways of N-bearing molecules in different elemental abundances.}

\vspace{-15pt}

\subsection{Complex Organic Molecules}\label{subsec4.5}
We found that the molecular abundances of \ce{CH3OH} and other oxygen-bearing COMs fell at time of around $10^4$ yrs for the model with $\zeta = 1.36 \times 10^{-16}$ s$^{-1}$ and did not fit the observations at around $10^5$ yrs (as shown in Figure~\ref{fig:5settozero}), unlike other nitrogen- and sulphur-bearing species. This discrepancy may be due to the fact that we ran the chemical reaction network calculations with physically stationary conditions. If the COMs mainly exist in the accretion flow towards the central protostar, they are continuously supplied tothe  gas phase from ice on dust grains when they pass their snowlines in the flow. If the accretion flow is fast enough to reach close to the central protostar within $\sim$$10^4$ yrs after passing the snowlines, we would be able to observe high abundances of COMs even around \mbox{$10^5$ yrs} (\mbox{e.g.,~\cite{Nomura2009}}).

 \ce{CH3OH} is the main reservoir of the larger organic complex molecules, such as dimethyl ether and methyl formate. These molecules are formed on the surface of dust grains too. In the gas phase, \ce{CH3OCH3} is totally produced from the dissociative recombination reaction of \ce{CH3OCH4+} with an electron, while \ce{HCOOCH3} is produced from the reaction of the protonated methyl formate (\ce{H5C2O2+}) with an electron. Both \ce{CH3OCH4+} and \ce{H5C2O2+} are formed from methanol. Since the abundances of these COMs depend on methanol's abundance, their resultant abundances have the same trend and do not fit with observations at $10^5$ yrs for a CRIR of $1.36 \times 10^{-16}$ s$^{-1}$. We may need to consider physically non-stationary conditions, such as accretion flow, in order to fit the observations of these~COMs.

In the gas phase, \ce{H2CO} is formed from the reaction of CH$_3$ with O, and it fits the observations well at all temperatures around $10^5$ years and fits the best at the lowest temperature of 50 K. It was found that formyl cation (\ce{HCO+}) was formed from the reaction of \ce{CO} with \ce{H3+}, and it participated in the formation of \ce{H3O+}, which is one of the main precursors in the formation of oxygen-bearing species. At the age of $10^5$ years and at a value of \mbox{$1.36 \times 10^{-16}$ s$^{-1}$} for CRIR, \ce{HCO+} theoretical abundance matches the observational abundance at relatively lower densities.

\section{Conclusions}\label{sec5}

We investigated the chemistry in LMC hot cores by deriving the initial molecular abundance in the low-metallicity LMC environment and ran time-dependent chemical reaction network calculations with varying temperatures, densities and CRIRs. Three different trends of the resultant molecular abundances were obtained. First, the abundances of parent molecules decreased with time. The other two categories included daughter molecules, where the abundances of one increased with time, and the other had abundances that increased and then decreased. The results show that the variation in CRIRs has the greatest impact on the molecular abundances. Although varying temperatures and densities affect the molecular abundances as well, their impacts are not very significant compared with~CRIR.

By comparing our results from model calculations with the ALMA observations of molecular lines towards LMC hot cores, ST16, N105–2A and N113 A1, we obtained the best-fit values at the core age of $10^5$ yrs and the CRIR of $1.36 \times 10^{-16}$ s$^{-1}$.
This comparison with observations of the S-bearing species revealed that SO \ce{SO2} and OCS fit at all temperatures, while CS and \ce{H2CS} fit the best at lower temperatures. In the case of N-bearing species, a comparison between the model results and observations showed that NO, HNCO and CN fit at all temperatures. \ce{HC3N} fits the best at higher temperatures.
Moving on to the COMs, their resultant abundances did not fit with observations, which may suggest that COMs mainly exist in physically non-stationary conditions, for example, in accretion flow. The rotation temperature is assumed or derived when obtaining the abundance of each molecule from the observations in~\cite{Shimon2020}. We compared the temperatures with those at which our model fit the observed molecular abundances, and as a result, we found that these temperatures were consistent with each other for S- and N-bearing species.

To improve the models of the LMC hot cores, more observations are needed to reveal further information about the physical conditions in LMC hot cores, especially for CRIR. In addition to this, chemical models that include networks with more complex organic molecules are needed to investigate the formation of astrobiological molecules in low-metallicity environments in the LMC.

\vspace{6pt} 




\authorcontributions{Conceptualisation, Y.S. and H.N.; methodology, Y.S.; software, Y.S. and H.N.; validation, H.N., O.S. and T.Y.; formal analysis, Y.S.; investigation, Y.S.; resources, H.N.; data curation, Y.S.; writing---original draft preparation, Y.S.; writing---review and editing, H.N.; visualisation, Y.S.; supervision, H.N.; project administration, H.N. and O.S.; funding acquisition, H.N. All authors have read and agreed to the published version of the manuscript.}

\funding{HN acknowledges the support given by JSPS and MEXT Grants-in-Aid for Scientific Research, 18H05441, 19K03910, and 20H00182.}

\dataavailability{The original code used for the chemical reaction network calculations in this paper is available at UMIST Database {website} \url{http://www.udfa.net} that was accessed on 1 June 2021. The data underlying this article will be shared on reasonable request
to the corresponding author.}  



\acknowledgments{We would like to thank Takashi Shimonishi and Kenji Furuya for their useful~\mbox{discussions}. We would like to thank the reviewers for their useful comments.}

\conflictsofinterest{The authors declare no conflicts of interest.}

\begin{adjustwidth}{-\extralength}{0cm}
 
\setenotez{list-name=Note}
\printendnotes[custom] 

\reftitle{References}

\PublishersNote{}
\end{adjustwidth}

\begin{thebibliography}{999}

\bibitem[Blake et~al.(1987)Blake, Sutton, Masson, and Phillips]{Blake1987}
Blake, G.A.; Sutton, E.; Masson, C.; Phillips, T.
\newblock Molecular abundances in OMC-1: The chemical composition of
  interstellar molecular clouds and the influence of massive star formation.
\newblock {\em {Astrophys. J.}} {\bf 1987}, {\em 315},~621--645. 

\bibitem[{Wright} and {Plambeck}(2017)]{Wright2017}
{Wright}, M.C.H.; {Plambeck}, R.L.
\newblock {ALMA Images of the Orion Hot Core at 349 GHz}.
\newblock {\em Astrophys. J.} {\bf 2017}, {\em 843},~83.

\bibitem[{Shimonishi} et~al.(2016){Shimonishi}, {Onaka}, {Kawamura}, and
  {Aikawa}]{Shimon2016}
{Shimonishi}, T.; {Onaka}, T.; {Kawamura}, A.; {Aikawa}, Y.
\newblock {The Detection of a Hot Molecular Core in the Large Magellanic Cloud
  with ALMA}.
\newblock {\em Astrophys. J.} {\bf 2016}, {\em 827},~72.

\bibitem[{Pietrzy{\'n}ski} et~al.(2013){Pietrzy{\'n}ski}, {Graczyk}, {Gieren},
  {Thompson}, {Pilecki}, {Udalski}, {Soszy{\'n}ski}, {Koz{\l}owski},
  {Konorski}, {Suchomska}, {Bono}, {Moroni}, {Villanova}, {Nardetto},
  {Bresolin}, {Kudritzki}, {Storm}, {Gallenne}, {Smolec}, {Minniti}, {Kubiak},
  {Szyma{\'n}ski}, {Poleski}, {Wyrzykowski}, {Ulaczyk}, {Pietrukowicz},
  {G{\'o}rski}, and {Karczmarek}]{Pietrz2013}
{Pietrzy{\'n}ski}, G.; {Graczyk}, D.; {Gieren}, W.; {Thompson}, I.B.;
  {Pilecki}, B.; {Udalski}, A.; {Soszy{\'n}ski}, I.; {Koz{\l}owski}, S.;
  {Konorski}, P.; {Suchomska}, K.;  et~al.
\newblock {An eclipsing-binary distance to the Large Magellanic Cloud accurate
  to two per cent}.
\newblock {\em Nature} {\bf 2013}, {\em 495},~76--79.

\bibitem[{Russell} and {Dopita}(1992)]{Russell1992}
{Russell}, S.C.; {Dopita}, M.A.
\newblock {Abundances of the Heavy Elements in the Magellanic Clouds. III.
  Interpretation of Results}.
\newblock {\em Astrophys. J.} {\bf 1992}, {\em 384},~508.

\bibitem[{Dufour} et~al.(1982){Dufour}, {Shields}, and {Talbot}]{Dufour1982}
{Dufour}, R.J.; {Shields}, G.A.; {Talbot}, R.~J., J.
\newblock {The carbon abundance in the Magellanic Clouds from IUE
  observations.}
\newblock {\em Astrophys. J.} {\bf 1982}, {\em 252},~461--473.

\bibitem[{Westerlund}(1990)]{Wester1990}
{Westerlund}, B.E.
\newblock {The Magellanic Clouds: their evolution, structure and composition}.
\newblock {\em  Astron. Astrophys. Rev.} {\bf 1990}, {\em 2},~29--78.

\bibitem[{Rolleston} et~al.(2002){Rolleston}, {Trundle}, and
  {Dufton}]{Rolleston2002}
{Rolleston}, W.R.J.; {Trundle}, C.; {Dufton}, P.L.
\newblock {The present-day chemical composition of the LMC}.
\newblock {\em Astron. Astrophys.} {\bf 2002}, {\em 396},~53--64.

\bibitem[{Sewi{\l}o} et~al.(2018){Sewi{\l}o}, {Indebetouw}, {Charnley},
  {Zahorecz}, {Oliveira}, {van Loon}, {Ward}, {Chen}, {Wiseman}, {Fukui},
  {Kawamura}, {Meixner}, {Onishi}, and {Schilke}]{Sewilo2018}
{Sewi{\l}o}, M.; {Indebetouw}, R.; {Charnley}, S.B.; {Zahorecz}, S.;
  {Oliveira}, J.M.; {van Loon}, J.T.; {Ward}, J.L.; {Chen}, C.H.R.; {Wiseman},
  J.; {Fukui}, Y.;  et~al.
\newblock {The Detection of Hot Cores and Complex Organic Molecules in the
  Large Magellanic Cloud}.
\newblock {\em Astrophys. J. Lett.} {\bf 2018}, {\em 853},~L19.

\bibitem[{Shimonishi} et~al.(2020){Shimonishi}, {Das}, {Sakai}, {Tanaka},
  {Aikawa}, {Onaka}, {Watanabe}, and {Nishimura}]{Shimon2020}
{Shimonishi}, T.; {Das}, A.; {Sakai}, N.; {Tanaka}, K.E.I.; {Aikawa}, Y.;
  {Onaka}, T.; {Watanabe}, Y.; {Nishimura}, Y.
\newblock {Chemistry and Physics of a Low-metallicity Hot Core in the Large
  Magellanic Cloud}.
\newblock {\em Astrophys. J.} {\bf 2020}, {\em 891},~164.

\bibitem[{Sewi{\l}o} et~al.(2022){Sewi{\l}o}, {Cordiner}, {Charnley},
  {Oliveira}, {Garcia-Berrios}, {Schilke}, {Ward}, {Wiseman}, {Indebetouw},
  {Tokuda}, {van Loon}, {S{\'a}nchez-Monge}, {Allen}, {Chen}, {Hamedani
  Golshan}, {Karska}, {Kristensen}, {Kurtz}, {M{\"o}ller}, {Onishi}, and
  {Zahorecz}]{Sewilo2022}
{Sewi{\l}o}, M.; {Cordiner}, M.; {Charnley}, S.B.; {Oliveira}, J.M.;
  {Garcia-Berrios}, E.; {Schilke}, P.; {Ward}, J.L.; {Wiseman}, J.;
  {Indebetouw}, R.; {Tokuda}, K.;  et~al.
\newblock {ALMA Observations of Molecular Complexity in the Large Magellanic
  Cloud: The N 105 Star-forming Region}.
\newblock {\em Astrophys. J.} {\bf 2022}, {\em 931},~102.

\bibitem[{Acharyya} and {Herbst}(2015)]{Acharyya2015}
{Acharyya}, K.; {Herbst}, E.
\newblock {Molecular Development in the Large Magellanic Cloud}.
\newblock {\em Astrophys. J.} {\bf 2015}, {\em 812},~142.

\bibitem[{Shimonishi} et~al.(2016){Shimonishi}, {Dartois}, {Onaka}, and
  {Boulanger}]{Shimonb2016}
{Shimonishi}, T.; {Dartois}, E.; {Onaka}, T.; {Boulanger}, F.
\newblock {VLT/ISAAC infrared spectroscopy of embedded high-mass YSOs in the
  Large Magellanic Cloud: Methanol and the 3.47 {\ensuremath{\mu}}m band}.
\newblock {\em Astron. Astrophys.} {\bf 2016}, {\em 585},~A107.

\bibitem[{Nishimura} et~al.(2016){Nishimura}, {Shimonishi}, {Watanabe},
  {Sakai}, {Aikawa}, {Kawamura}, and {Yamamoto}]{Nishimura2016}
{Nishimura}, Y.; {Shimonishi}, T.; {Watanabe}, Y.; {Sakai}, N.; {Aikawa}, Y.;
  {Kawamura}, A.; {Yamamoto}, S.
\newblock {Spectral Line Survey toward Molecular Clouds in the Large Magellanic
  Cloud}.
\newblock {\em Astrophys. J.} {\bf 2016}, {\em 818},~161.

\bibitem[{Brown} et~al.(1988){Brown}, {Charnley}, and {Millar}]{Brown1988}
{Brown}, P.D.; {Charnley}, S.B.; {Millar}, T.J.
\newblock {A model of the chemistry in hot molecular cores.}
\newblock {\em Mon. Not. R. Astron. Soc.} {\bf 1988}, {\em 231},~409--417.

\bibitem[{Caselli} et~al.(1993){Caselli}, {Hasegawa}, and
  {Herbst}]{Caselli1993}
{Caselli}, P.; {Hasegawa}, T.I.; {Herbst}, E.
\newblock {Chemical Differentiation between Star-forming Regions: The Orion Hot
  Core and Compact Ridge}.
\newblock {\em Astrophys. J.} {\bf 1993}, {\em 408},~548.

\bibitem[{Millar} et~al.(1997){Millar}, {MacDonald}, and {Gibb}]{Millar1997}
{Millar}, T.J.; {MacDonald}, G.H.; {Gibb}, A.G.
\newblock {A 330-360 GHz spectral survey of G 34.3+0.15. II. Chemical
  modelling.}
\newblock {\em Astron. Astrophys.} {\bf 1997}, {\em 325},~1163--1173.

\bibitem[{Nomura} and {Millar}(2004)]{Nomura2004}
{Nomura}, H.; {Millar}, T.J.
\newblock {The physical and chemical structure of hot molecular cores}.
\newblock {\em Astron. Astrophys.} {\bf 2004}, {\em 414},~409--423.

\bibitem[{Millar} and {Herbst}(1990)]{Millar1990}
{Millar}, T.J.; {Herbst}, E.
\newblock {Chemical modelling of dark clouds in the LMC and SMC.}
\newblock {\em Mon. Not. R. Astron. Soc.} {\bf 1990}, {\em 242},~92--97.


\bibitem[{Pauly} and {Garrod}(2018)]{Pauly2018}
{Pauly}, T.; {Garrod}, R.T.
\newblock {Modelling CO, CO$_{2}$, and H$_{2}$O Ice Abundances in the Envelopes
  of Young Stellar Objects in the Magellanic Clouds}.
\newblock {\em Astrophys. J.} {\bf 2018}, {\em 854},~13.


\bibitem[{McElroy} et~al.(2013){McElroy}, {Walsh}, {Markwick}, {Cordiner},
  {Smith}, {Millar}]{Mc2013}
{McElroy}, D.; {Walsh}, C.; {Markwick}, A.~J.; {Cordiner}, M.~A.; {Smith}, K.; {Millar}, T.~J.
\newblock {The UMIST database for astrochemistry 2012}.
\newblock {\em Astron. Astrophys.} {\bf 2013}, {\em 550},~A36.
\newblock {\url{https://doi.org/10.1051/0004-6361/201220465}}.


\bibitem[{Hatchell} et~al.(1998){Hatchell}, {Thompson}, {Millar}, and
  {MacDonald}]{Hatchell1998}
{Hatchell}, J.; {Thompson}, M.A.; {Millar}, T.J.; {MacDonald}, G.H.
\newblock {Sulphur chemistry and evolution in hot cores}.
\newblock {\em Astron. Astrophys.} {\bf 1998}, {\em 338},~713--722.

\bibitem[{Padovani} et~al.(2024){Padovani}, {Galli}, {Scarlett}, {Grassi},
  {Rehill}, {Zammit}, {Bray}, and {Fursa}]{Padovani2024}
{Padovani}, M.; {Galli}, D.; {Scarlett}, L.H.; {Grassi}, T.; {Rehill}, U.S.;
  {Zammit}, M.C.; {Bray}, I.; {Fursa}, D.V.
\newblock {Ultraviolet H$_{2}$ luminescence in molecular clouds induced by
  cosmic rays}.
\newblock {\em Astron. Astrophys.} {\bf 2024}, {\em 682},~A131.
\newblock {\url{https://doi.org/10.1051/0004-6361/202348168}}.

\bibitem[{Abdo} et~al.(2010){Abdo}, {Ackermann}, {Ajello}, {Atwood}, {Baldini},
  {Ballet}, {Barbiellini}, {Bastieri}, {Baughman}, {Bechtol}, {Bellazzini},
  {Berenji}, {Blandford}, {Bloom}, {Bonamente}, {Borgland}, {Bregeon}, {Brez},
  {Brigida}, {Bruel}, {Burnett}, {Buson}, {Caliandro}, {Cameron}, {Caraveo},
  {Casandjian}, {Cecchi}, {{\c{C}}elik}, {Chekhtman}, {Cheung}, {Chiang},
  {Ciprini}, {Claus}, {Cohen-Tanugi}, {Cominsky}, {Conrad}, {Cutini}, {Dermer},
  {de Angelis}, {de Palma}, {Digel}, {Silva}, {Drell}, {Dubois}, {Dumora},
  {Farnier}, {Favuzzi}, {Fegan}, {Focke}, {Fortin}, {Frailis}, {Fukazawa},
  {Fusco}, {Gargano}, {Gasparrini}, {Gehrels}, {Germani}, {Giavitto},
  {Giebels}, {Giglietto}, {Giordano}, {Glanzman}, {Godfrey}, {Gotthelf},
  {Grenier}, {Grondin}, {Grove}, {Guillemot}, {Guiriec}, {Hanabata}, {Harding},
  {Hayashida}, {Hays}, {Horan}, {Hughes}, {Jackson}, {Jean}, {J{\'o}hannesson},
  {Johnson}, {Johnson}, {Johnson}, {Johnson}, {Kamae}, {Katagiri}, {Kataoka},
  {Kawai}, {Kerr}, {Kn{\"o}dlseder}, {Kocian}, {Kuss}, {Lande}, {Latronico},
  {Lemoine-Goumard}, {Longo}, {Loparco}, {Lott}, {Lovellette}, {Lubrano},
  {Madejski}, {Makeev}, {Marshall}, {Martin}, {Mazziotta}, {McConville},
  {McEnery}, {Meurer}, {Michelson}, {Mitthumsiri}, {Mizuno}, {Moiseev},
  {Monte}, {Monzani}, {Morselli}, {Moskalenko}, {Murgia}, {Nolan}, {Norris},
  {Nuss}, {Ohsugi}, {Omodei}, {Orlando}, {Ormes}, {Paneque}, {Parent},
  {Pelassa}, {Pepe}, {Pesce-Rollins}, {Piron}, {Porter}, {Rain{\`o}}, {Rando},
  {Razzano}, {Reimer}, {Reimer}, {Reposeur}, {Ritz}, {Rodriguez}, {Romani},
  {Roth}, {Ryde}, {Sadrozinski}, {Sanchez}, {Sander}, {Saz Parkinson},
  {Scargle}, {Sellerholm}, {Sgr{\`o}}, {Siskind}, {Smith}, {Smith}, {Spandre},
  {Spinelli}, {Starck}, {Strickman}, {Strong}, {Suson}, {Tajima}, {Takahashi},
  {Tanaka}, {Thayer}, {Thayer}, {Thompson}, {Tibaldo}, {Torres}, {Tosti},
  {Tramacere}, {Uchiyama}, {Usher}, {Vasileiou}, {Venter}, {Vilchez}, {Vitale},
  {Waite}, {Wang}, {Weltevrede}, {Winer}, {Wood}, {Ylinen}, and
  {Ziegler}]{Abdo2010}
{Abdo}, A.A.; {Ackermann}, M.; {Ajello}, M.; {Atwood}, W.B.; {Baldini}, L.;
  {Ballet}, J.; {Barbiellini}, G.; {Bastieri}, D.; {Baughman}, B.M.; {Bechtol},
  K.;  et~al.
\newblock {Observations of the Large Magellanic Cloud with Fermi}.
\newblock {\em Astron. Astrophys.} {\bf 2010}, {\em 512},~A7.

\bibitem[{Maggi} et~al.(2016){Maggi}, {Haberl}, {Kavanagh}, {Sasaki},
  {Bozzetto}, {Filipovi{\'c}}, {Vasilopoulos}, {Pietsch}, {Points}, {Chu},
  {Dickel}, {Ehle}, {Williams}, and {Greiner}]{Maggi2016}
{Maggi}, P.; {Haberl}, F.; {Kavanagh}, P.J.; {Sasaki}, M.; {Bozzetto}, L.M.;
  {Filipovi{\'c}}, M.D.; {Vasilopoulos}, G.; {Pietsch}, W.; {Points}, S.D.;
  {Chu}, Y.H.;  et~al.
\newblock {The population of X-ray supernova remnants in the Large Magellanic
  Cloud}.
\newblock {\em Astron. Astrophys.} {\bf 2016}, {\em 585},~A162.

\bibitem[{Ou} et~al.(2018){Ou}, {Chu}, {Maggi}, {Li}, {Chang}, and
  {Gruendl}]{Ou2018}
{Ou}, P.S.; {Chu}, Y.H.; {Maggi}, P.; {Li}, C.J.; {Chang}, U.P.; {Gruendl},
  R.A.
\newblock {X-ray Luminosity and Size Relationship of Supernova Remnants in the
  LMC}.
\newblock {\em Astrophys. J.} {\bf 2018}, {\em 863},~137.

\bibitem[Vaupr{\'e} et~al.(2014)Vaupr{\'e}, Hily-Blant, Ceccarelli, Dubus,
  Gabici, and Montmerle]{Vaupr2014}
Vaupr{\'e}, S.; Hily-Blant, P.; Ceccarelli, C.; Dubus, G.; Gabici, S.;
  Montmerle, T.
\newblock Cosmic ray induced ionisation of a molecular cloud shocked by the W28
  supernova remnant.
\newblock {\em Astron. Astrophys.} {\bf 2014}, {\em 568},~A50.

\bibitem[{Morales Ortiz} et~al.(2014){Morales Ortiz}, {Ceccarelli}, {Lis},
  {Olmi}, {Plume}, and {Schilke}]{Morales2014}
{Morales Ortiz}, J.L.; {Ceccarelli}, C.; {Lis}, D.C.; {Olmi}, L.; {Plume}, R.;
  {Schilke}, P.
\newblock {Ionization toward the high-mass star-forming region NGC 6334 I}.
\newblock {\em Astron. Astrophys.} {\bf 2014}, {\em 563},~A127.

\bibitem[{Cabedo} et~al.(2023){Cabedo}, {Maury}, {Girart}, {Padovani},
  {Hennebelle}, {Houde}, and {Zhang}]{Cabedo2023}
{Cabedo}, V.; {Maury}, A.; {Girart}, J.M.; {Padovani}, M.; {Hennebelle}, P.;
  {Houde}, M.; {Zhang}, Q.
\newblock {Magnetically regulated collapse in the B335 protostar?. II.
  Observational constraints on gas ionization and magnetic field coupling}.
\newblock {\em Astron. Astrophys.} {\bf 2023}, {\em 669},~A90.

\bibitem[Asplund et~al.(2021)Asplund, Amarsi, and Grevesse]{Asplund2021}
Asplund, M.; Amarsi, A.; Grevesse, N.
\newblock The chemical make-up of the Sun: A 2020 vision.
\newblock {\em Astron. Astrophys.} {\bf 2021}, {\em 653},~A141.

\bibitem[{Garrod}(2013)]{Garrod2013}
{Garrod}, R.T.
\newblock {A Three-phase Chemical Model of Hot Cores: The Formation of
  Glycine}.
\newblock {\em Astrophys. J.} {\bf 2013}, {\em 765},~60. 
\newblock {\url{https://doi.org/10.1088/0004-637X/765/1/60}}.

\bibitem[{Walsh} et~al.(2014){Walsh}, {Millar}, {Nomura}, {Herbst}, {Widicus
  Weaver}, {Aikawa}, {Laas}, and {Vasyunin}]{Walsh2014}
{Walsh}, C.; {Millar}, T.J.; {Nomura}, H.; {Herbst}, E.; {Widicus Weaver}, S.;
  {Aikawa}, Y.; {Laas}, J.C.; {Vasyunin}, A.I.
\newblock {Complex organic molecules in protoplanetary disks}.
\newblock {\em Astron. Astrophys.} {\bf 2014}, {\em 563},~A33.

\bibitem[{Entekhabi} et~al.(2022){Entekhabi}, {Tan}, {Cosentino}, {Hsu},
  {Caselli}, {Walsh}, {Lim}, {Henshaw}, {Barnes}, {Fontani}, and
  {Jim{\'e}nez-Serra}]{Entekhabi2022}
{Entekhabi}, N.; {Tan}, J.C.; {Cosentino}, G.; {Hsu}, C.J.; {Caselli}, P.;
  {Walsh}, C.; {Lim}, W.; {Henshaw}, J.D.; {Barnes}, A.T.; {Fontani}, F.;
  et~al.
\newblock {Astrochemical modelling of infrared dark clouds}.
\newblock {\em Astron. Astrophys.} {\bf 2022}, {\em 662},~A39.

\bibitem[{Yorke}(2004)]{York2004}
{Yorke}, H.W.
\newblock {Theory of Formation of Massive Stars via Accretion}.
\newblock In \emph{Proceedings of the Star Formation at High Angular Resolution}; 
  {Burton}, M.G., {Jayawardhana}, R., {Bourke}, T.L., Eds.; {Astronomical Society of the Pacific: San Francis, CA, USA},  2004; IAU Symposium 221, p. 141. 

\bibitem[{Herbst} and {van Dishoeck}(2009)]{Herbst2009}
{Herbst}, E.; {van Dishoeck}, E.F.
\newblock {Complex Organic Interstellar Molecules}.
\newblock {\em Annu. Rev. Astron. Astrophys.} {\bf 2009}, {\em 47},~427--480.

\bibitem[{Charnley} et~al.(1992){Charnley}, {Tielens}, and
  {Millar}]{Charnley1992}
{Charnley}, S.B.; {Tielens}, A.G.G.M.; {Millar}, T.J.
\newblock {On the Molecular Complexity of the Hot Cores in Orion A: Grain
  Surface Chemistry as ``The Last Refuge of the Scoundrel''}.
\newblock {\em Astrophys. J. Lett.} {\bf 1992}, {\em 399},~L71.  

\bibitem[{Wilner} et~al.(2001){Wilner}, {De Pree}, {Welch}, and
  {Goss}]{Wilner2001}
{Wilner}, D.J.; {De Pree}, C.G.; {Welch}, W.J.; {Goss}, W.M.
\newblock {Hot Cores in W49N and the Timescale for Hot Core Evolution}.
\newblock {\em Astrophys. J. Lett.} {\bf 2001}, {\em 550},~L81--L85.

\bibitem[{Charnley}(1997)]{Charnley1997}
{Charnley}, S.B.
\newblock {Sulfuretted Molecules in Hot Cores}.
\newblock {\em Astrophys. J.} {\bf 1997}, {\em 481},~396--405.

\bibitem[{Rodgers} and {Charnley}(2001)]{Rodgers2001}
{Rodgers}, S.D.; {Charnley}, S.B.
\newblock {Chemical Differentiation in Regions of Massive Star Formation}.
\newblock {\em Astrophys. J.} {\bf 2001}, {\em 546},~324--329.

\bibitem[{Nomura} et~al.(2009){Nomura}, {Aikawa}, {Nakagawa}, and
  {Millar}]{Nomura2009}
{Nomura}, H.; {Aikawa}, Y.; {Nakagawa}, Y.; {Millar}, T.J.
\newblock {Effects of accretion flow on the chemical structure in the inner
  regions of protoplanetary disks}.
\newblock {\em Astron. Astrophys.} {\bf 2009}, {\em 495},~183--188.

\end{thebibliography}
\end{document}